\pdfoutput=1

\documentclass[12pt,a4paper]{article}

\usepackage{multirow}
\usepackage{ifthen} 
\newboolean{pdflatex}
\setboolean{pdflatex}{true} 

\newboolean{articletitles}
\setboolean{articletitles}{true} 

\newboolean{uprightparticles}
\setboolean{uprightparticles}{false} 

\usepackage{caption} 

\newboolean{inbibliography}
\setboolean{inbibliography}{false} 

\usepackage{microtype}
\usepackage{lineno}  
\usepackage{xspace} 
\usepackage{caption} 

\usepackage{graphicx}  
\usepackage{color}
\usepackage{colortbl}
\graphicspath{{./figs/}} 

\usepackage{amsmath} 
\usepackage{amssymb}
\usepackage{amsfonts}
\usepackage{upgreek} 

\newcommand*\patchAmsMathEnvironmentForLineno[1]{%
\expandafter\let\csname old#1\expandafter\endcsname\csname #1\endcsname
\expandafter\let\csname oldend#1\expandafter\endcsname\csname
end#1\endcsname
 \renewenvironment{#1}%
   {\linenomath\csname old#1\endcsname}%
   {\csname oldend#1\endcsname\endlinenomath}%
}
\newcommand*\patchBothAmsMathEnvironmentsForLineno[1]{%
  \patchAmsMathEnvironmentForLineno{#1}%
  \patchAmsMathEnvironmentForLineno{#1*}%
}
\AtBeginDocument{%
\patchBothAmsMathEnvironmentsForLineno{equation}%
\patchBothAmsMathEnvironmentsForLineno{align}%
\patchBothAmsMathEnvironmentsForLineno{flalign}%
\patchBothAmsMathEnvironmentsForLineno{alignat}%
\patchBothAmsMathEnvironmentsForLineno{gather}%
\patchBothAmsMathEnvironmentsForLineno{multline}%
\patchBothAmsMathEnvironmentsForLineno{eqnarray}%
}

\usepackage{hyperref}    
\usepackage[all]{hypcap} 

\def\lhcb {\mbox{LHCb}\xspace}

\def\MagUp {\mbox{\em Mag\kern -0.05em Up}\xspace}

\ifthenelse{\boolean{uprightparticles}}%
{

 \def\Pmu         {\ensuremath{\upmu}\xspace}

 \def\Ppi         {\ensuremath{\uppi}\xspace}

 \def\Ppsi        {\ensuremath{\uppsi}\xspace}

 \def\PDelta      {\ensuremath{\Delta}\xspace}                 
 \def\PXi      {\ensuremath{\Xi}\xspace}                 
 \def\PLambda      {\ensuremath{\Lambda}\xspace}                 
 \def\PSigma      {\ensuremath{\Sigma}\xspace}                 
 \def\POmega      {\ensuremath{\Omega}\xspace}                 
 \def\PUpsilon      {\ensuremath{\Upsilon}\xspace}                 
 
 \def\PB      {\ensuremath{\mathrm{B}}\xspace}                 
                  
 \def\PD      {\ensuremath{\mathrm{D}}\xspace}

 \def\PJ      {\ensuremath{\mathrm{J}}\xspace}                 
 \def\PK      {\ensuremath{\mathrm{K}}\xspace}

 \def\Pb      {\ensuremath{\mathrm{b}}\xspace}                 
 \def\Pc      {\ensuremath{\mathrm{c}}\xspace}                 
 \def\Pd      {\ensuremath{\mathrm{d}}\xspace}

 \def\Pi      {\ensuremath{\mathrm{i}}\xspace}

 \def\Pp      {\ensuremath{\mathrm{p}}\xspace}

 \def\Ps      {\ensuremath{\mathrm{s}}\xspace}

}
{

 \def\Pmu         {\ensuremath{\mu}\xspace}

 \def\Ppi         {\ensuremath{\pi}\xspace}

 \def\Ppsi        {\ensuremath{\psi}\xspace}                 
                  
 \mathchardef\PDelta="7101
 \mathchardef\PXi="7104
 \mathchardef\PLambda="7103
 \mathchardef\PSigma="7106
 \mathchardef\POmega="710A
 \mathchardef\PUpsilon="7107
                  
 \def\PB      {\ensuremath{B}\xspace}                 
                  
 \def\PD      {\ensuremath{D}\xspace}

 \def\PJ      {\ensuremath{J}\xspace}                 
 \def\PK      {\ensuremath{K}\xspace}

 \def\Pb      {\ensuremath{b}\xspace}                 
 \def\Pc      {\ensuremath{c}\xspace}                 
 \def\Pd      {\ensuremath{d}\xspace}

 \def\Pi      {\ensuremath{i}\xspace}

 \def\Pp      {\ensuremath{p}\xspace}

 \def\Ps      {\ensuremath{s}\xspace}

}

\makeatletter
\ifcase \@ptsize \relax
  \newcommand{\miniscule}{\@setfontsize\miniscule{4}{5}}
\or
  \newcommand{\miniscule}{\@setfontsize\miniscule{5}{6}}
\or
  \newcommand{\miniscule}{\@setfontsize\miniscule{5}{6}}
\fi
\makeatother

\DeclareRobustCommand{\optbar}[1]{\shortstack{{\miniscule (\rule[.5ex]{1.25em}{.18mm})}
  \\ [-.7ex] $#1$}}

\def\muon       {{\ensuremath{\Pmu}}\xspace}

\def\dquark    {{\ensuremath{\Pd}}\xspace}

\def\squark    {{\ensuremath{\Ps}}\xspace}

\def\cquark    {{\ensuremath{\Pc}}\xspace}

\def\bquark    {{\ensuremath{\Pb}}\xspace}

\def\pion   {{\ensuremath{\Ppi}}\xspace}

\def\pip    {{\ensuremath{\pion^+}}\xspace}
\def\pim    {{\ensuremath{\pion^-}}\xspace}

\def\kaon    {{\ensuremath{\PK}}\xspace}
  \def\Kbar    {{\kern 0.2em\overline{\kern -0.2em \PK}{}}\xspace}

\def\KorKbar    {\kern 0.18em\optbar{\kern -0.18em K}{}\xspace}

\def\Km      {{\ensuremath{\kaon^-}}\xspace}

\def\KS      {{\ensuremath{\kaon^0_{\rm\scriptscriptstyle S}}}\xspace}

  \def\Dbar    {{\kern 0.2em\overline{\kern -0.2em \PD}{}}\xspace}

\def\DorDbar    {\kern 0.18em\optbar{\kern -0.18em D}{}\xspace}

\def\B       {{\ensuremath{\PB}}\xspace}
\def\Bbar    {{\ensuremath{\kern 0.18em\overline{\kern -0.18em \PB}{}}}\xspace}

\def\BorBbar    {\kern 0.18em\optbar{\kern -0.18em B}{}\xspace}
\def\Bz      {{\ensuremath{\B^0}}\xspace}
\def\Bzb     {{\ensuremath{\Bbar{}^0}}\xspace}

\def\Bd      {{\ensuremath{\B^0}}\xspace}
\def\Bs      {{\ensuremath{\B^0_\squark}}\xspace}
\def\Bsb     {{\ensuremath{\Bbar{}^0_\squark}}\xspace}
\def\Bdb     {{\ensuremath{\Bbar{}^0}}\xspace}

\def\jpsi     {{\ensuremath{{\PJ\mskip -3mu/\mskip -2mu\Ppsi\mskip 2mu}}}\xspace}

  \def\Y#1S{\ensuremath{\PUpsilon{(#1S)}}\xspace}

\def\proton      {{\ensuremath{\Pp}}\xspace}

\def\Lz          {{\ensuremath{\PLambda}}\xspace}
\def\Lbar        {{\ensuremath{\kern 0.1em\overline{\kern -0.1em\PLambda}}}\xspace}
\def\LorLbar    {\kern 0.18em\optbar{\kern -0.18em \PLambda}{}\xspace}
\def\porpbar    {\kern 0.18em\optbar{\kern -0.18em \proton}{}\xspace}

\def\Lb      {{\ensuremath{\Lz^0_\bquark}}\xspace}

\def\to                 {\ensuremath{\rightarrow}\xspace}

\def\CP                {{\ensuremath{C\!P}}\xspace}

\newcommand{\dmd}{{\ensuremath{\Delta m_{\dquark}}}\xspace}

\newcommand{\DGd}{{\ensuremath{\Delta\Gamma_{\dquark}}}\xspace}

\newcommand{\Gd}{{\ensuremath{\Gamma_{\dquark}}}\xspace}

\def\AT#1     {\ensuremath{A_{\mathrm{T}}^{#1}}\xspace}           

\def\C#1      {\ensuremath{\mathcal{C}_{#1}}\xspace}                       
\def\Cp#1     {\ensuremath{\mathcal{C}_{#1}^{'}}\xspace}                    
\def\Ceff#1   {\ensuremath{\mathcal{C}_{#1}^{\mathrm{(eff)}}}\xspace}        
\def\Cpeff#1  {\ensuremath{\mathcal{C}_{#1}^{'\mathrm{(eff)}}}\xspace}       
\def\Ope#1    {\ensuremath{\mathcal{O}_{#1}}\xspace}                       
\def\Opep#1   {\ensuremath{\mathcal{O}_{#1}^{'}}\xspace}                    

\newcommand{\tev}{\ifthenelse{\boolean{inbibliography}}{\ensuremath{~T\kern -0.05em eV}\xspace}{\ensuremath{\mathrm{\,Te\kern -0.1em V}}}\xspace}
\newcommand{\gev}{\ensuremath{\mathrm{\,Ge\kern -0.1em V}}\xspace}
\newcommand{\mev}{\ensuremath{\mathrm{\,Me\kern -0.1em V}}\xspace}
\newcommand{\kev}{\ensuremath{\mathrm{\,ke\kern -0.1em V}}\xspace}
\newcommand{\ev}{\ensuremath{\mathrm{\,e\kern -0.1em V}}\xspace}
\newcommand{\gevc}{\ensuremath{{\mathrm{\,Ge\kern -0.1em V\!/}c}}\xspace}
\newcommand{\mevc}{\ensuremath{{\mathrm{\,Me\kern -0.1em V\!/}c}}\xspace}
\newcommand{\gevcc}{\ensuremath{{\mathrm{\,Ge\kern -0.1em V\!/}c^2}}\xspace}
\newcommand{\gevgevcccc}{\ensuremath{{\mathrm{\,Ge\kern -0.1em V^2\!/}c^4}}\xspace}
\newcommand{\mevcc}{\ensuremath{{\mathrm{\,Me\kern -0.1em V\!/}c^2}}\xspace}

\def\m    {\ensuremath{\rm \,m}\xspace}

\def\mum  {\ensuremath{{\,\upmu\rm m}}\xspace}

\def\gsim{{~\raise.15em\hbox{$>$}\kern-.85em
          \lower.35em\hbox{$\sim$}~}\xspace}
\def\lsim{{~\raise.15em\hbox{$<$}\kern-.85em
          \lower.35em\hbox{$\sim$}~}\xspace}

\newcommand{\Imag}{\ensuremath{\mathcal{I}m}\xspace}

\def\ptot       {\mbox{$p$}\xspace}
\def\pt         {\mbox{$p_{\rm T}$}\xspace}

\def\evtgen     {\mbox{\textsc{EvtGen}}\xspace}

\def\gauss      {\mbox{\textsc{Gauss}}\xspace}
\def\geant      {\mbox{\textsc{Geant4}}\xspace}

\def\photos     {\mbox{\textsc{Photos}}\xspace}

\def\pythia     {\mbox{\textsc{Pythia}}\xspace}

\def\tell1  {TELL1\xspace}
\def\ukl1   {UKL1\xspace}

\usepackage{epsfig,accents}
\newcommand*{\fancybar}{\scalebox{.4}{(}\raisebox{-1.7pt}{--}\scalebox{.4}{)}}
\newcommand*{\brabar}[1]{\accentset{\fancybar}{#1}}

\def \m {m_{hh}}
\def \angmu {\theta_{\jpsi}}
\def \angpi {\theta_{hh}}

\def \Ab {\overline{A}}
\def \cs {\cos(\dmd t)}
\def \sn {\sin(\dmd t)}

\def \A {{\cal A}}
\def \cAb {\overline{{\cal A}}}

\DeclareRobustCommand{\optbar}[1]{\shortstack{{\miniscule (\rule[.5ex]{1.25em}{.18mm})}
  \\ [-.7ex] $#1$}}

\usepackage{mathrsfs}
\usepackage{cite} 
\usepackage{mciteplus}

\usepackage{longtable} 

\begin{document}

\renewcommand{\thefootnote}{\fnsymbol{footnote}}
\setcounter{footnote}{1}



\begin{titlepage}
\pagenumbering{roman}

\vspace*{-1.5cm}
\centerline{\large EUROPEAN ORGANIZATION FOR NUCLEAR RESEARCH (CERN)}
\vspace*{1.5cm}
\hspace*{-0.5cm}
\begin{tabular*}{\linewidth}{lc@{\extracolsep{\fill}}r}
\ifthenelse{\boolean{pdflatex}}
{\vspace*{-2.7cm}\mbox{\!\!\!\includegraphics[width=.14\textwidth]{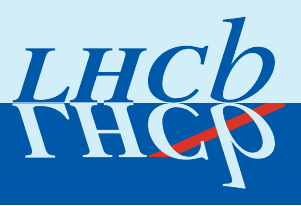}} & &}%
{\vspace*{-1.2cm}\mbox{\!\!\!\includegraphics[width=.12\textwidth]{lhcb-logo.eps}} & &}%
\\
 & & CERN-PH-EP-2014-268 \\  
 & & LHCb-PAPER-2014-058 \\  
 & & \today \\ 
 & & \\
\end{tabular*}

\vspace*{4.0cm}

{\bf\boldmath\huge
\begin{center}
 Measurement of the \CP-violating phase $\beta$ in $\Bz\rightarrow J/\psi \pi^+\pi^-$ decays and limits on penguin effects
\end{center}
}

\vspace*{1.5cm}

\begin{center}
The LHCb collaboration\footnote{Authors are listed at the end of this letter.}
\end{center}

\vspace{\fill}

\begin{abstract}
  \noindent
 Time-dependent \CP violation is measured in the $\BorBbar^0\to\jpsi\pi^+\pi^-$ channel for each $\pi^+\pi^-$ resonant final state using data collected with an integrated luminosity of 3.0\,fb$^{-1}$ in $pp$ collisions using the LHCb detector. The final state with the largest rate, $\jpsi\rho^0(770)$, is used to measure the \CP-violating angle $2\beta^{\rm eff}$ to be $(41.7\pm9.6_{-6.3}^{+2.8})^{\circ}$.  This result can be used to limit the size of penguin amplitude contributions to \CP violation measurements in, for example,
 $\BorBbar^0_{\!\!s}\to \jpsi\phi$ decays. Assuming approximate SU(3) flavour symmetry and neglecting higher order diagrams, the shift in the \CP-violating phase $\phi_s$ is limited to be within the interval [$-1.05^\circ$, +$1.18^\circ$] at 95\% confidence level. Changes to the limit due to SU(3) symmetry breaking effects are also discussed.
\end{abstract}

\vspace*{1.5cm}

\begin{center}
  Submitted to Phys.~Lett.~B
 \end{center}

\vspace{\fill}

{\footnotesize
\centerline{\copyright~CERN on behalf of the \lhcb collaboration, license \href{http://creativecommons.org/licenses/by/4.0/}{CC-BY-4.0}.}}
\vspace*{2mm}

\end{titlepage}


\newpage
\setcounter{page}{2}
\mbox{~}
%

%
%
%

\cleardoublepage


\renewcommand{\thefootnote}{\arabic{footnote}}
\setcounter{footnote}{0}



\pagestyle{plain} 
\setcounter{page}{1}
\pagenumbering{arabic}


%

\section{Introduction}
\begin{figure}[htb]
\vskip -.2cm
\begin{center}
\includegraphics[width=6in]{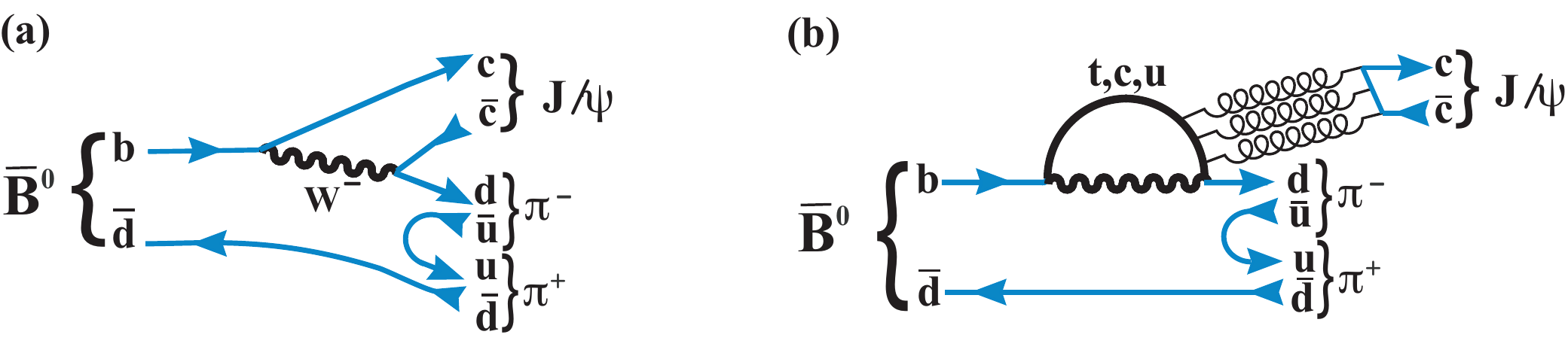}
\end{center}
\vskip -.5cm
\caption{\small (a) Tree level and (b) penguin diagram for $\Bzb$ decays into $J/\psi \pi^+\pi^-$.}\label{feyn3}
\end{figure}

Measurements of \CP violation in neutral $B$ meson decays are used either to search for physics beyond the Standard Model (SM) \cite{Glashow:1961tr,*Weinberg:1967tq} or set limits on combinations of Cabibbo-Kobayashi-Maskawa couplings $(V_{ij})$ \cite{Kobayashi:1973fv}. Interpretations of the measurement of the \CP-violating phase $2\beta$ via the interference of mixing and decays in the $\BorBbar^0\to\jpsi\KS$ channel, and the phase $\phi_s$ in $\BorBbar^0_{\!\!s}\to\jpsi\phi$ and $\jpsi\pi^+\pi^-$ decays,\footnote{\CP violation measurements in $\BorBbar^0\to\jpsi\KS$ determine the sum of  $2\beta\equiv 2\arg(-V_{cd}V_{cb}^*)/(V_{td}V_{tb}^*)$ and contributions from higher order diagrams. Similar measurements in the $\BorBbar^0_{\!\!s}$ system determine  $\phi_s$ which is the sum of $-2\beta_s\equiv -2\arg(-V_{ts}V_{tb}^*)/(V_{cs}V_{cb}^*)$ and higher order corrections.}
are made assuming that the decays are dominated by tree-level processes. However penguin processes are also possible, and they may have large enough amplitudes to influence the results. Here we use $\BorBbar^0\to\jpsi\pi^+\pi^-$ decays to set limits on possible changes due to penguin contributions. This mode has both tree and penguin diagrams, as shown in
Fig.~\ref{feyn3}.
  Theoretical models, to be discussed later, predict that the ratio of penguin to tree amplitudes is greatly enhanced in this decay relative to $\BorBbar^0\to\jpsi \KS$ \cite{Fleischer:1999sj,Faller:2008zc,*Faller:2008gt}. Thus, the effects of penguin topologies can be investigated  by using the $J/\psi \pi^+ \pi^-$ decay  and comparing different measurements of the \CP-violating phase $2\beta$  in $\jpsi \KS$, and individual channels such as $\BorBbar^0 \to J/\psi \rho^0(770)$.\footnote{In the following $\rho^0$ or $\rho$ refers to the $\rho^0(770)$ meson.}

Next, we discuss the time-dependent decay rate, taking into account that the $\pi^+\pi^-$ system is composed of the resonances previously reported in Ref.~\cite{Aaij:2014siy}. This analysis largely follows the measurement procedure used in the study of \CP violation in $\BorBbar^0_{\!\!s}$ decays into $J/\psi \pi^+\pi^-$ \cite{Aaij:2014dka}.
The total decay amplitude for $\BorBbar^0$ at a decay time of zero is taken to be the sum over individual $\pi^+\pi^-$ resonant transversity amplitudes \cite{Dighe:1995pd}, and possibly one non-resonant amplitude, with each component labelled as ${A}_i$ ($\overline{A}_i$). The quantities $q$ and $p$ relate the mass eigenstates to the flavour eigenstates \cite{Bigi:2000yz}. By introducing the parameter $\lambda_i \equiv \frac{q}{p}\frac{\Ab_i}{A_i}$, relating \CP violation in the interference between mixing and decay associated with the state $i$, the amplitudes ${\cal A}$ and $\cAb$ can be expressed as the sums of the individual $\BorBbar^0$ amplitudes,  ${\cal A}=\sum A_i$ and $\cAb =\sum \frac{q}{p} \Ab_i =\sum \lambda_i A_i= \sum \eta_i |\lambda_i|e^{-i2\beta_i^{\rm eff}}A_i$. For each transversity state $i$ the \CP-violating phase $2\beta_i^{\rm eff}\equiv -\arg(\eta_i\lambda_i)$ with $\eta_i$ being the \CP eigenvalue of the state.\footnote{Note that while ${q}/{p}$ and ${\Ab_i}/{A_i}$ are phase convention dependent, $\lambda_i$ is not.}
The decay rates are\footnote{We assume $\DGd=0$ and $|p/q|=1$. The averages of current measurements are  $\Delta\Gamma_d/\Gamma_d=0.001\pm 0.010$ and $|p/q|=1.0005\pm 0.0011$ \cite{HFAG}.}
\begin{equation*}\label{Eq-ts}
\Gamma(t) =
  {\cal N} e^{-\Gd t}\left\{\frac{|\A|^2+|\cAb|^2}{2}  + \frac{|\A|^2-|\cAb|^2}{2}\cs  -  \Imag(\A^*\cAb)\sn\right\},~~~~
\end{equation*}
\begin{equation}\label{Eqbar-ts}
\overline{\Gamma}(t) =
 {\cal N}  e^{-\Gd t}\left\{\frac{|\A|^2+|\cAb|^2}{2}  - \frac{|\A|^2-|\cAb|^2}{2}\cs+  \Imag(\A^*\cAb)\sn\right\}.
 \end{equation}

\section{Penguin and tree amplitudes}

The decay $\Bz\to\jpsi\KS$ can be written as the sum of one tree level amplitude, similar to that shown in Fig.~\ref{feyn3}(a), but where the virtual $W^-$ transforms to a $\overline{c}s$ pair, and three penguin amplitudes similar to those shown in Fig.~\ref{feyn3}(b). Here we neglect higher order diagrams. The $t$-quark mediated penguin amplitude can be expressed in terms of the other two using CKM unitarity. The resulting decay amplitude is \cite{Fleischer:1999sj}
\begin{equation}
A(\Bz\to\jpsi\KS)=\left(1-\frac{\widetilde{\lambda}^2}{2}\right){\mathscr{A}}\left[1+\frac{\widetilde{\lambda}^2}{1-\widetilde{\lambda}^2}ae^{i\theta}e^{i\gamma}\right],
\label{eq:AjpsiK}
\end{equation}
where $\widetilde{\lambda}=|V_{us}|=0.2252$ \cite{PDG}, $\gamma\equiv \arg\left( -{V_{ud}V_{ub}^*}/{V_{cd}V_{cb}^*}\right)$, ${\mathscr{A}}$ denotes the sum of tree and penguin strong amplitudes, and $a$ and $\theta$ are the magnitude and phase of the strong parts of the effective penguin amplitude relative to the tree amplitude.

For the case of  $\BorBbar^0\to\jpsi\pi^+\pi^-$ decays, the $\pi^+\pi^-$ pairs are in spin states ranging from zero to two. Since they are in a final state with a spin-1 \jpsi resonance, the amplitudes in the different transversity states $f$ need to be distinguished for all spins above zero. For example, the amplitude for each $\jpsi\rho^0(770)$ transversity state is\begin{equation}
-\sqrt{2}A(\Bz\to(\jpsi\rho)_f)=\widetilde{\lambda}{\mathscr{A'}}\left[1-a_f'e^{i\theta_f'}e^{i\gamma}\right],
\end{equation}
where the primed quantities are defined in analogy with the unprimed ones in Eq.~(\ref{eq:AjpsiK}). For \Bzb decays only the sign in front of $i\gamma$ changes. We are only concerned here with the relative size of the tree and penguin amplitudes. For $\jpsi\KS$ the penguin is suppressed relative to the tree by an additional factor of $\epsilon\equiv
\widetilde{\lambda}^2/({1-\widetilde{\lambda}^2})=0.0534$. Thus,  comparing even a relatively poor measurement of $2\beta^{\rm eff}$ measured in $\jpsi\rho^0$  with $2\beta$ measured in $\jpsi\KS$ allows us to set stringent limits on the penguin contribution. Using approximate SU(3) flavour symmetry the size of the penguin contribution in $\BorBbar^0\to\jpsi\rho^0$  can be related to that in $\BorBbar^0_{\!\!s}\to\jpsi\phi$ decays as pointed out in Refs.~\cite{Faller:2008zc,*Faller:2008gt,Fleischer:1999zi}.

We now turn to the expressions for \CP violation in the presence of both tree and penguin amplitudes. The  complex-valued \CP parameter $\lambda_f$ is given by
\begin{equation}
\lambda_f\equiv\frac{q}{p}\frac{A(\Bdb \to (\jpsi \rho)_f)}{A(\Bd \to (\jpsi \rho)_f)}=\eta_f \frac{1-a_f^\prime e^{i\theta_f^\prime}e^{-i\gamma}}{1-a_f^\prime e^{i\theta_f^\prime}e^{i\gamma}}e^{-2i\beta},
\end{equation}
where $\beta$ is the phase induced by mixing. Thus the measured phase $\beta^{\rm eff}_f$ is related to $\beta$ by
\begin{equation}\label{pen1}
\eta_f \lambda_f\equiv |\lambda_f| e^{-i2\beta^{\rm eff}_{\!f}} =\frac{1-a_f^\prime e^{i\theta_f^\prime}e^{-i\gamma}}{1-a_f^\prime e^{i\theta_f^\prime}e^{i\gamma}}e^{-i2\beta},
\end{equation}
separating real and imaginary parts gives
\begin{equation}
\label{Eq17}
|\lambda_f|=\left|\frac{1-a_f^\prime e^{i\theta_f^\prime}e^{-i\gamma}}{1-a_f^\prime e^{i\theta_f^\prime}e^{i\gamma}}\right|,{\rm ~and~}
\Delta2\beta_f\equiv 2\beta_f^{\rm eff}-2\beta = -\arg\left(\frac{1-a_f^\prime e^{i\theta_f^\prime}e^{-i\gamma}}{1-a_f^\prime e^{i\theta_f^\prime}e^{i\gamma}}\right).
\end{equation}

For the $\jpsi\KS$ mode we replace $a'_f$ and $\theta'_f$ in Eq.~(\ref{Eq17}) by $-\epsilon a_f$ and $\theta_f$, respectively.  In addition, we take $a'=a$ and $\theta'=\theta$. The relationship between the penguin influence on the mixing induced \CP violation phase in favoured decays and the measurements in $\BorBbar^0\to(\jpsi \rho^0)_f$ is then given by
\begin{equation}\label{EqRel}
\delta_P=-\arg\left(\frac{(\lambda^\prime_f e^{2i\gamma}-1)+\epsilon(\lambda^\prime_f-1)}{(\lambda^\prime_f e^{2i\gamma}-1)+\epsilon(\lambda^\prime_f-1)e^{2i\gamma}}\right)~{\rm where~} \lambda_f^\prime\equiv|\lambda_f| e^{-i\Delta 2\beta_f}.
\end{equation}
We will show that the penguin shift has a weak dependence on $|\lambda_f|$, resulting in 
$\delta_P\approx-\epsilon\Delta 2\beta_f$.
Since the uncertainty on the current measurement of 2$\beta$ is $(^{+1.6}_{-1.5})^{\circ}$, a measurement of $\Delta 2\beta_f$, even with an uncertainty ten times larger, could limit penguin contributions to be well below the current statistical uncertainty, which is the main aim of this analysis.

\section{Detector software and event selection}
\label{sec:Detector}

The \lhcb detector~\cite{Alves:2008zz} is a single-arm forward
spectrometer covering the \mbox{pseudorapidity} range $2<\eta <5$,
designed for the study of particles containing \bquark or \cquark
quarks. The detector includes a high-precision tracking system
consisting of a silicon-strip vertex detector surrounding the $pp$
interaction region~\cite{LHCb-DP-2014-001}, a large-area silicon-strip detector located
upstream of a dipole magnet with a bending power of about
$4{\rm\,Tm}$, and three stations of silicon-strip detectors and straw
drift tubes~\cite{LHCb-DP-2013-003} placed downstream of the magnet.
The tracking system provides a measurement of momentum\footnote{We use natural units where $\hbar$=c=1.}, \ptot,  with
a relative uncertainty that varies from 0.4\% at low momentum to 0.6\% at 100\,GeV.
The minimum distance of a track to a primary vertex (PV), the impact parameter (IP), is measured with a resolution of $(15+29/\pt)\mum$,
where \pt is the component of \ptot transverse to the beam, in GeV.
Different types of charged hadrons are distinguished using information
from two ring-imaging Cherenkov detectors~\cite{LHCb-DP-2012-003}. Photon, electron and
hadron candidates are identified by a calorimeter system consisting of
scintillating-pad and preshower detectors, an electromagnetic
calorimeter and a hadronic calorimeter. Muons are identified by a
system composed of alternating layers of iron and multiwire
proportional chambers~\cite{LHCb-DP-2012-002}.

The trigger~\cite{Aaij:2012me} consists of a hardware stage, based
on information from the calorimeter and muon systems, followed by a
software stage that applies a full event reconstruction~\cite{Aaij:2012me}. Events selected for this analysis are triggered by a $J/\psi\to\mu^+\mu^-$  decay, where the
$J/\psi$ meson is required at the software level to be consistent with coming from the decay of a $\BorBbar^0$ meson by use of either of IP requirements or detachment of the $J/\psi$ meson decay vertex  from the primary vertex.
In the simulation, $pp$ collisions are generated using
\pythia~\cite{Sjostrand:2006za,*Sjostrand:2007gs}
 with a specific \lhcb
configuration~\cite{LHCb-PROC-2010-056}.  Decays of hadronic particles
are described by \evtgen~\cite{Lange:2001uf}, in which final state
radiation is generated using \photos~\cite{Golonka:2005pn}. The
interaction of the generated particles with the detector and its
response are implemented using the \geant
toolkit~\cite{Allison:2006ve, *Agostinelli:2002hh} as described in
Ref.~\cite{LHCb-PROC-2011-006}.

A $\BorBbar^0 \to \jpsi \pi^+\pi^-$ candidate is reconstructed by combining a $\jpsi \to \mu^+\mu^-$ candidate with two pions of opposite charge. The like-sign combinations $\jpsi\pi^\pm\pi^\pm$ are also reconstructed for background studies. 
 The event selection is described in detail in the time-integrated amplitude analysis~\cite{Aaij:2014siy}.  The only difference here is that we reject $\KS\to\pi^+\pi^-$ candidates by excluding the events in the region within $\pm$20\,MeV of the $\KS$ mass peak.

Only candidates with dimuon invariant mass between $-$48\mev and +43\mev relative to the observed $\jpsi$ mass peak are selected, corresponding a window of  about $\pm3\sigma$. The two muons subsequently are kinematically constrained to the known $\jpsi$ mass.
Other requirements are imposed to isolate $\Bzb$ candidates with high signal yield and minimum background. This is accomplished by combining the $\jpsi\to\mu^+\mu^-$ candidate with a pair of pion candidates of opposite charge, and then testing if all four tracks form a common decay vertex.
Pion candidates are each required to have $\pt$ greater than 250\mev, and the scalar sum of the two transverse momenta, $\pt(\pi^+)+\pt(\pi^-)$, must be larger than 900\mev. To test for inconsistency with production at the PV, 
the IP $\chi^2$ is computed as the difference between the $\chi^2$ of the PV reconstructed with and without the considered track. Each pion must have an IP $\chi^2$ greater than 9. 
Pion and kaon candidates are positively identified using the RICH system.
The four-track \Bzb candidate must have a flight distance of more than 1.5~mm, where the average decay length resolution is 0.17~mm. The
angle between the combined momentum vector of the decay products
and the vector formed from the positions of the PV and
the decay vertex (pointing angle) is required to be less than $2.5^{\circ}$.
Events satisfying this preselection are then further filtered using a multivariate analyzer
based on a Boosted Decision Tree (BDT) technique~\cite{Breiman}. The BDT uses eight variables that
are chosen to provide separation between signal and background.
These are the minimum of DLL($\mu-\pi$) of the $\mu^+$ and $\mu^-$, $\pt(\pi^+)+\pt(\pi^-)$, the minimum of IP $\chi^2$ of the $\pi^+$ and $\pi^-$, and the $\Bzb$ properties of vertex $\chi^2$, pointing angle, flight distance, $\pt$ and IP $\chi^2$, where DLL($\mu-\pi$) is a logarithm of the likelihood
ratio between $\muon$ and $\pion$ hypotheses for the muon candidates.

The BDT is trained on a simulated sample of two million $\Bzb\to \jpsi \pi^+\pi^-$ signal events generated uniformly in phase space with unpolarized $J/\psi \rightarrow \mu^+\mu^-$ decays, and a background data sample from the sideband $5566<m(J/\psi \pi^+\pi^-)< 5616$\,\mev. Then separate samples are used to train and test the BDT. 
\begin{figure}[t]
\begin{center}
\vskip -.5cm
\includegraphics[scale=0.55]{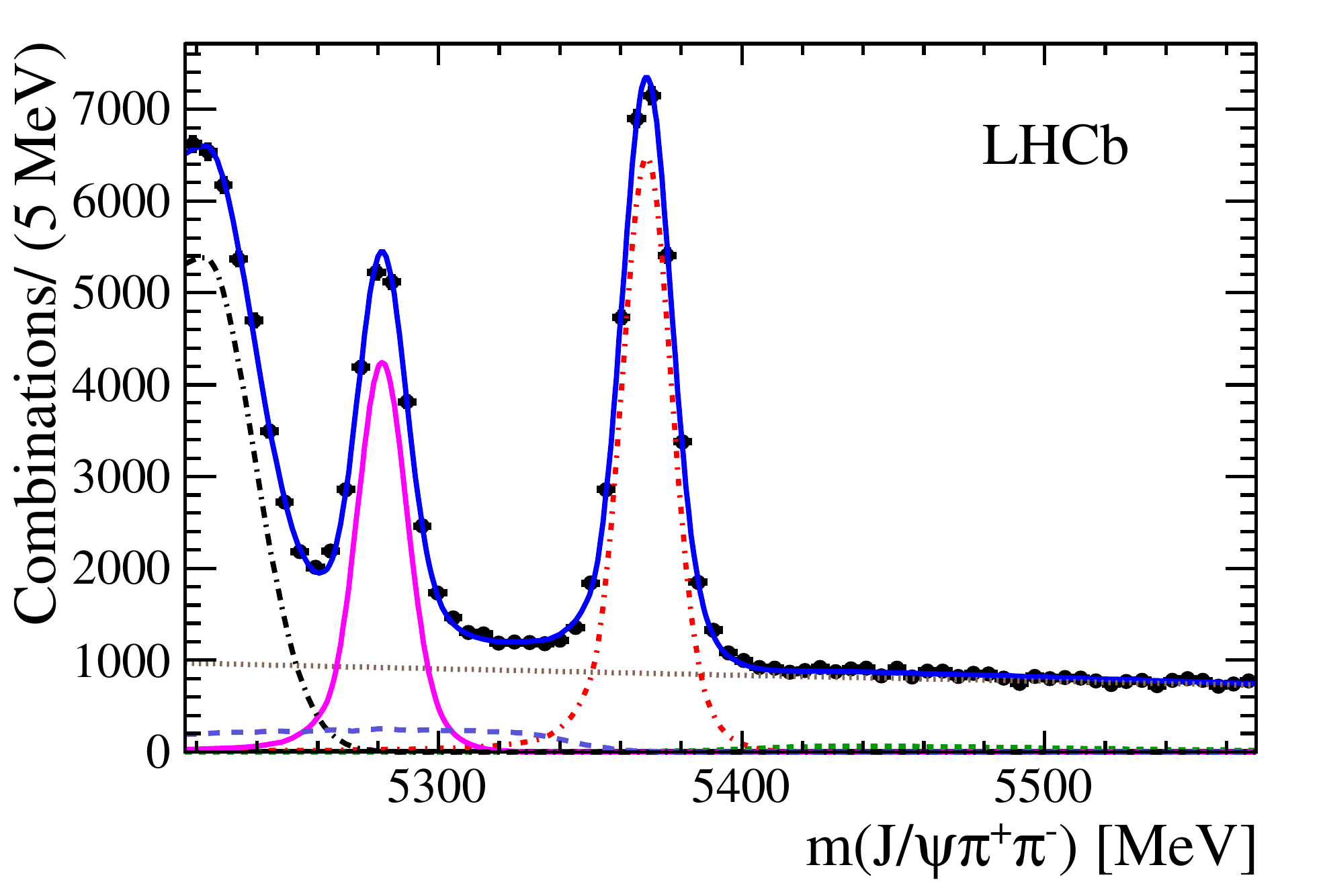}
\end{center}
\vskip -0.9cm
\caption{Invariant mass of $J/\psi \pi^+\pi^-$ combinations with $\KS$ veto. The data have been fitted with double-Crystal ball signal and several background functions. The (purple) solid line shows the $\Bzb$ signal, the (brown) dotted line shows the combinatorial background, the (green) short-dashed shows the $B^-$ background, the (red) dot-dashed  is $\Bsb\rightarrow J/\psi \pi^+\pi^-$, the (light blue) long-dashed is the sum of $\Bsb\rightarrow J/\psi\eta'$, $\Bsb\rightarrow J/\psi\phi$ when  $\phi\to\pi^+\pi^-\pi^0$ backgrounds and the $\Lb \to \jpsi K^- p$ reflection, the (black) dot-long dashed is the $\Bdb\rightarrow J/\psi K^- \pi^+$ reflection and the (blue) solid line is the total.}\label{fitmass}
\end{figure}

The invariant mass of the selected $\jpsi\pi^+\pi^-$ combinations, where the dimuon pair is
constrained to have the $\jpsi$  mass, is shown in Fig.~\ref{fitmass}.
There is a large peak at the $\Bsb$ mass and a smaller one at the $\Bzb$ mass on top of the background. A double Crystal Ball function with common means models the radiative tails and is used to fit each of the  signals \cite{Skwarnicki:1986xj}. Other components in the fit model take into account background contributions from  $B^-\rightarrow J/\psi K^-$ and $B^-\rightarrow J/\psi \pi^-$ decays combined with a random $\pi^+$, $\Bsb\rightarrow J/\psi\eta(')$ with $\eta(')\rightarrow \pi^+\pi^- \gamma$, $\Bsb\rightarrow J/\psi\phi$  with $\phi\rightarrow \pi^+\pi^-\pi^0$,  $\Bdb\rightarrow J/\psi K^- \pi^+$ and $\Lb \to \jpsi \Km \proton$ reflections, and combinatorial backgrounds. The exponential  combinatorial background shape is taken from like-sign combinations, that are the sum of $\pi^+\pi^+$ and $\pi^-\pi^-$ candidates. The shapes of the other components are taken
from the simulation with their normalizations allowed to vary. Only the candidates within
$\pm20$ MeV of the $\Bz$ mass peak are retained for CP violation measurements; the fit gives
$17\,650 \pm 200$ signal and $9\,840 \pm 160$ background candidates.

\section{The signal likelihood}

We fit the entire $\pi^+\pi^-$ mass spectrum, by including the resonance contributions found in the amplitude analysis \cite{Aaij:2014siy}, in order to measure the  \CP-violating parameters of all the states, the most important being $\BorBbar^0\to\jpsi\rho^0$ as it has the largest fit fraction of approximately 65\%.  The same likelihood construction as was used to determine the \CP-violating quantities $\phi_s$ and $|\lambda|$ in $\BorBbar^0_{\!\!s}\to\jpsi\pi^+\pi^-$ decays \cite{Aaij:2014dka} is employed. Here the value of $\Delta\Gamma_d\approx 0$ simplifies some terms, and the smaller value of $\Delta m_d$ makes the decay time resolution function less important.  In addition, a different same-sign flavour tagging algorithm is used.

The determination of the \CP violation parameters relies upon the formalism developed in Ref.~\cite{Zhang:2012zk}.
For $\jpsi$ decays to $\mu^+\mu^-$ final states  the amplitudes are themselves functions of four variables: the $\pi^+\pi^-$ invariant mass $\m = m(\pi^+\pi^-)$,
and three angles $\Omega$, defined in the helicity basis. These consist of: $\angmu$,
the angle between the $\mu^+$ direction in the $\jpsi$ rest frame with respect to the $\jpsi$ direction in the $\BorBbar^0$ rest frame;  $\angpi$, the angle between the $h^+$ direction in the $h^+h^-$ rest frame with respect to the $h^+h^-$ direction in the $\BorBbar^0$ rest frame;
and $\chi$,  the angle between the $\jpsi$ and $h^+h^-$ decay planes in the $\BorBbar^0$ rest frame \cite{Zhang:2012zk,Aaij:2013oba}.

\begin{figure}[b]
\begin{center}
    \includegraphics[width=0.8\textwidth]{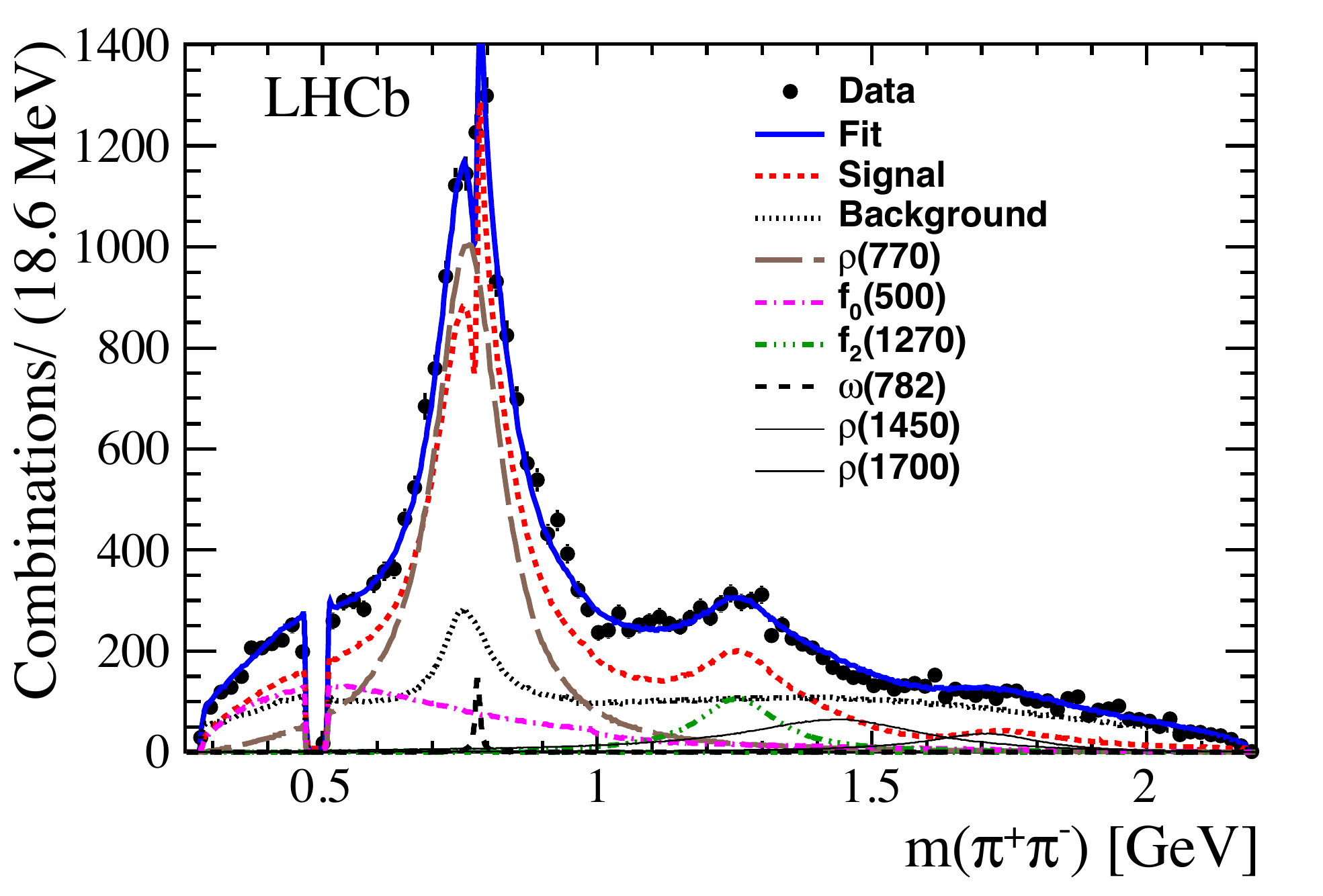}%
\end{center}
\vskip -0.6cm
\caption{\small Fit projection of $m(\pi^+\pi^-)$ showing the different resonant contributions in the ``Best model" \cite{Aaij:2014siy}. The \KS veto causes the absence of events near 500\,MeV. The shape variation near 780\,MeV is due to interference between the $\rho(770)$ and $\omega(782)$ states. The total fit is the sum of the individual components plus their interferences.}
\label{B0-mpp}
\end{figure}

We perform a simultaneous unbinned maximum likelihood fit to the decay time $t$,  $m_{hh}$, and the three helicity angles $\Omega$, along with information on the initial flavour of the decaying hadron, $i.e.$ whether it was produced as a \Bz or a \Bzb meson. The probability density function (PDF) used in the fit consists of signal and background components that include detector resolution and acceptance effects.  The predicted decay time error for each event is used for the decay time resolution model, and similarly the measured per-event misidentification probability is used for determining the initial flavour of the neutral $B$ meson.
The $\pi^+\pi^-$ invariant mass distribution
is shown in  Fig.~\ref{B0-mpp} along with the fitted components of the different resonances using the ``Best model" \cite{Aaij:2014siy} for the $\pi^+\pi^-$ resonance content.

Knowledge of the $\BorBbar^0$ flavour at production, called ``tagging", is necessary to measure \CP violation.  We use both opposite-side (OS)~\cite{LHCb-PAPER-2011-027} and same-side pion  (SS$\pi$) tagging information; here we use the same procedure as for same-side kaon tagging used in the $\BorBbar^0_{\!\!s}\to \jpsi \pi^+\pi^-$ and $\jpsi\phi$ analyses  \cite{Aaij:2013oba}, but identify the tag from a pion rather than a kaon. The wrong-tag probability $\eta$ is estimated  based on the output of a neural network trained on simulated data. It is calibrated with data using flavour-specific decay modes in order to predict the true wrong-tag probability of the event $\brabar{\omega}(\eta)$ for an initial flavour $\BorBbar^0$ meson, which has a linear dependence on $\eta$.
The calibration is performed separately for the OS and the SS$\pi$ taggers. 
If events are tagged by both OS and SS$\pi$ algorithms, a combined tag decision and wrong-tag probability are given by the algorithm defined in Ref.~\cite{LHCb-PAPER-2011-027}. This  combined algorithm is implemented in the overall fit. The effective tagging power obtained is characterized by $\varepsilon_{\rm tag}D^2=(3.26\pm 0.17)\%$, where $D\equiv(1-2\omega_{\rm avg})$ is the dilution, $\omega_{\rm avg}$ is the average wrong-tag probability for $\omega$ and $\bar{\omega}$, and $\varepsilon_{\rm tag}=(42.1\pm0.6)\%$ is  the signal tagging efficiency.

The signal decay time distribution including flavour tagging is

\begin{align}\label{eq:R}
R(\hat{t},\m,\Omega,\mathfrak{q}|\eta) =&\frac{1}{1+|\mathfrak{q}|}\bigg[\left[1+\mathfrak{q}\left(1-2\omega(\eta)\right)\right]\Gamma(\hat{t},\m,\Omega)\bigg.\nonumber\\
&\bigg. +\left[1-\mathfrak{q}\left(1-2\bar{\omega}(\eta)\right)\right]\frac{1+A_{\rm P}}{1-A_{\rm P}}\bar{\Gamma}(\hat{t},\m,\Omega)\bigg],
\end{align}
where $\hat{t}$ is the true decay time, $\brabar{\Gamma}$ is defined in Eq.~(\ref{Eqbar-ts}), and $A_{\rm P}=-0.0035\pm 0.0081$~\cite{Aaij:2014bba,*Aaij:2014nxa} is the $\Bzb-\Bz$ production asymmetry in the LHCb acceptance. The flavour tag parameter $\mathfrak{q}$ takes values of $-1$ or +1 if the signal meson is tagged as $\Bzb$, $\Bz$ respectively, or 0 if untagged.

The signal function is convolved with the decay time resolution and multiplied by the acceptance:
\begin{equation}
F^{\rm sig}(t,\m,\Omega,\mathfrak{q}|\eta,\delta_t)= \left[R(\hat{t},\m,\Omega,\mathfrak{q}|\eta) \otimes T(t-\hat{t};\delta_t)\right] \cdot {\cal E}_t(t) \cdot \varepsilon(\m,\Omega),
\end{equation}
where $\varepsilon(\m,\Omega)$ is the efficiency as a function of the $h^+h^-$ mass and angles, obtained from the simulation as described in Ref.~\cite{Aaij:2014siy}, $T(t-\hat{t};\delta_t)$ is the decay time resolution function which depends upon the estimated decay time error for each event $\delta_t$, and ${\cal E}_t(t)$ is the decay time acceptance function.
The decay time resolution function $T(t-\hat{t};\delta_t)$ is described by a sum of three Gaussian functions with a common mean. Studies using simulated data show that  $\jpsi \pi^+\pi^-$ combinations produced directly in the $pp$ interaction (prompt) have nearly identical resolution to signal events.
Specifically, the time resolution is determined using prompt $J/\psi$ decays into a dimuon pair, using a dedicated trigger for calibration purposes, plus two oppositely charged tracks from the primary vertex with the similar selection criteria as for $J/\psi \pi^+\pi^-$ and an invariant mass within $\pm$20 MeV of the $\Bz$ mass. The effective resolution is found to be about 40\,fs by using the weighted average widths of the three Gaussians. This is negligibly small compared to the \Bzb-\Bz oscillation time.

The decay time distribution is influenced by acceptance effects that are introduced by track reconstruction,  trigger and  event selection. The decay time acceptance is obtained using control samples of $\BorBbar^0 \to J/\psi \KorKbar^{*0}(\to K^{\mp}\pi^{\pm})$ decays, corrected by the acceptance ratio between $\jpsi K^{\mp}\pi^{\pm}$ and $\jpsi\pip\pim$ derived from simulation.

The acceptance function for the control sample is defined as
\begin{equation}
A(t;a,n,t_0,\beta_1,\beta_2) = \frac{\left[a(t-t_0)\right]^n}{1+\left[a(t-t_0)\right]^n} \times (1+\beta_1 t+\beta_2 t^2),
\end{equation}
where $a$, $n$, $t_0$, $\beta_1$, $\beta_2$ are parameters determined by the fit. The decay time distribution of $\BorBbar^0\to J/\psi K^\mp\pi^\pm$ candidates is described by the function
\begin{align}
P^{0}(t) =& \left(f_{\rm 0}A(t;a,n,t_0,\beta_1,\beta_2)\frac{e^{-\hat{t}/\tau_{\Bz}}}{\tau_{\Bz}{\cal N}_{\Bz}}+(1-f_{\rm 0})A(t;a_{\rm bkg}^{0},n_{\rm bkg}^{0},0,0,0)\frac{e^{-\hat{t}/\tau^0_{\rm bkg}}}{\tau^0_{\rm bkg}{\cal N}^0_{\rm bkg}}\right)\nonumber\\
&\otimes T(t-\hat{t};\delta_t),
\label{eq:acceptfun}
\end{align}
where $f_0$ is the signal fraction, and ${\cal N}_{\Bz}$ and ${\cal N}^0_{\rm bkg}$ are normalizations necessary to construct PDFs of signal and background, respectively. The background acceptance function in Eq.~(\ref{eq:acceptfun}) uses the same form as the signal and its parameters $a_{\rm bkg}^{0}$, $n_{\rm bkg}^{0}$ and $\tau^0_{\rm bkg}$ are obtained from mass sideband regions of 5180$-$5205\mev and 5400$-$5425\mev. The lifetime is constrained to $\tau_{\Bz}=1.519\pm0.007$ ps~\cite{PDG}. 

We use the product of the acceptance $A(a,n,t_0,\beta_1,\beta_2)$ determined from $\BorBbar^0 \to J/\psi \KorKbar^{*0}$ and the correction ratio found from simulation as the time acceptance function for $\BorBbar^0\to\jpsi\pi^+\pi^-$ events,
\begin{equation}\label{Eq:acc-final}
{\cal E}_t(t;a,n,t_0,\beta_1,\beta_2,p_1,p_2) = \frac{\left[a(t-t_0)\right]^n}{1+\left[a(t-t_0)\right]^n} \times (1+\beta_1 t+\beta_2 t^2)\times (1-p_2 e^{-p_1 t}),
\end{equation}
with parameter values and correlations given in Table \ref{acc-par}. 
\setlength{\tabcolsep}{5pt}
\begin{table}[hb]
\caption{Parameter values  and correlations for the acceptance function $\varepsilon_t(t)$ in Eq.~(\ref{Eq:acc-final}).}
\centering
\begin{tabular}{lrrrrrrr|l}\hline\hline
P&$n$~~ & $a$~~ & $\beta_1$~~ & $\beta_2$~~ & $t_0$~~ & $p_1$~~ & $p_2$~~ &~~~~~~~ Values\\\hline
$n$ &~~1.000 & ~~0.444 & ~~0.574& $-$0.536 &$-$0.862& ~~0.000 & ~~0.000 & ~~$2.082\pm0.036$\\
 $a$ &    &~~1.000 &~~0.739 &$-$0.735 &$-$0.050 & ~~0.000 & ~~0.000 & ~~$1.981\pm0.024$\,ps$^{-1}$\\
$\beta_1$  & &  & ~~1.000 &$-$0.899 &$-$0.374& ~~0.000 & ~~0.000 & ~~$0.077\pm0.009$\,ps$^{-1}$\\
$\beta_2$ && & & ~~1.000&  ~~0.343& ~~0.000 & ~~0.000 & $\!-0.008\pm0.001$\,ps$^{-2}$\\
 $t_0$ && && & ~~1.000& ~~0.000 & ~~0.000 & ~~$0.104\pm0.003$\,ps\\
  $p_1$ &&  & & &  & ~~1.000 & $-$0.885 & ~~$6.237\pm1.669$\,ps$^{-1}$ \\
$p_2$ &&&&&& & ~~1.000 & $\!-0.739\pm0.424$\\\hline\hline
\end{tabular}\label{acc-par}
\end{table}
\setlength{\tabcolsep}{6pt}

\section{\boldmath Measurements of 2$\beta^{\rm eff}$}

The \CP-violating parameters are determined from a fit that uses the amplitude model with six final state $\pi^+\pi^-$ resonances. In our previous amplitude analysis \cite{Aaij:2014siy} we used two parameterizations of the $f_0(500)$ resonance, ``default" and ``alternate". The default used a Breit-Wigner resonance shape, with relatively poorly measured parameters, while the alternate used a function suggested by Bugg \cite{Bugg:2006gc}, with more theoretically motivated shape parameters. In this analysis we choose to switch to the shape suggested by Bugg, while the Breit-Wigner shape of the previous default parameterization is used to assess systematic uncertainties.
 A Gaussian constraint using $\Delta m_d=0.510\pm0.003$\,ps$^{-1}$~\cite{PDG} is applied in the fit. All other parameters, such as the time resolution, and those describing the tagging are fixed. In addition to the \CP-violating parameters, the other free parameters are the amplitudes and phases of the resonances.
To minimize correlations in the fitted results, we choose as free parameters the \CP asymmetry $\alpha^i_{CP} = \frac{1-|\lambda_i|}{1+|\lambda_i|}$, $2\beta_{i}^{\rm eff}$ of the largest polarization component, and $\Delta 2\beta^{\rm eff}_{i}$ of the other components with respect to the largest one.

As $\jpsi\rho$ is the final state with the largest contribution, we treat it specially and perform two fits. In both cases all resonances other than the $\rho$ share a common \CP violation parameter $\lambda'$. For Fit 1 the three  $\rho$ transversity states share the same \CP violation parameter $\lambda$, while for  Fit 2 each  $\rho$ transversity state has its own \CP violation parameter $\lambda_i$.
 The results are shown in Table~\ref{datafit}. The statistical uncertainties are within $\pm15\%$ of the precision estimated using toy Monte Carlo simulation. To determine $\Delta2\beta_f$ we use the measured value in $b\to c \overline{c} s$ transitions of $(42.8^{+1.6}_{-1.5})^{\circ}$ found in $\BorBbar^0$ decays \cite{HFAG}. Our measurement of $2\beta^{\rm eff}$ is consistent with this value for both  Fit 1 and  Fit 2.
The correlation between $\alpha_{\CP}^{\rho}$ and $2\beta_{\rho}^{\rm eff}$ is $-0.01$ in Fit 1. Table~\ref{comatr} shows the correlation matrix for the \CP-violating parameters in Fit 2.
\begin{table}[b]
\begin{center}
\caption{Fit results for $2\beta_{i}^{\rm eff}$ and  $\alpha_{\CP}^i$.}\label{datafit}
\def\arraystretch{1.25}
\begin{tabular}{llc|lc}\hline
Condition& \multicolumn{2}{c|}{$2\beta_i^{\rm eff}$ ($^\circ$)}& \multicolumn{2}{c}{$\alpha_{\CP}^i (\times10^{-3})$}\\\hline
\multirow{2}{*}{Fit 1} &$\rho$ & $~~41.7\pm9.6_{-6.3}^{+2.8}$& $\rho$& $-32\pm28_{-~7}^{+~9}$\\
& other$-\rho$& $~~~~3.6\pm3.6_{-0.8}^{+0.9}$&other&$~\!-1\pm25_{-14}^{+~7}$\\\hline
& $\rho_0$& $~~44.1\pm10.2_{-6.9}^{+3.0}$ & $\rho_0$ & $\,-47\pm34_{-10}^{+11}$\\
\multirow{2}{*}{Fit 2}& $\rho_{\|}-\rho_0$& $-0.8\pm6.5_{-1.3}^{+1.9}$& $\rho_{\|}$& $~-61\pm60_{-~6}^{+~8}~~$\\
&$\rho_{\perp}-\rho_0$& $-3.6\pm7.2_{-1.4}^{+2.0}$& $\rho_{\perp}$& $~~~~17\pm109_{-15}^{+22}$\\
& other$-\rho_0$& $~~2.7\pm3.9_{-0.9}^{+1.0}$&other&$~~~~6\pm27_{-14}^{+~9}$\\
\hline
\end{tabular}
\def\arraystretch{1.0}
\end{center}
\end{table}

\begin{table}[b!]
\vspace{-2mm}
\begin{center}
\def\arraystretch{1.25}
\caption{The correlation matrix for the \CP-violating parameters determined using Fit 2, where $\Delta 2\beta^{\rm eff}_i=2\beta^{\rm eff}_i-2\beta^{\rm eff}_{\rho_0}$.\label{comatr} }
\begin{tabular}{ccccccccc}\hline
&$\alpha_{\CP}^{\rm other}$& $\alpha_{\CP}^{\rho_0}$ &$\alpha_{\CP}^{\rho_{\perp}}$& $\alpha_{\CP}^{\rho_{\|}}$ & $\Delta2\beta^{\rm eff}_{\rm other}$&$\Delta2\beta^{\rm eff}_{\rho_{\perp}}$&$\Delta2\beta^{\rm eff}_{\rho_\|}$&$2\beta^{\rm eff}_{\rho_0}$\\\hline

$\alpha_{\CP}^{\rm other}$	&$~~	1.00	$&$	-0.62	$&$	-0.28	$&$	-0.13	$&$~~	0.05	$&$	-0.42	$&$	-0.19	$&$~~	0.05	$	\\
$\alpha_{\CP}^{\rho_0}$ 	&		&$~~	1.00	$&$~~	0.03	$&$~~	0.16	$&$~~	0.29	$&$~~	0.22	$&$~~	0.16	$&$	-0.11	$	\\
$\alpha_{\CP}^{\rho_{\perp}}$	&		&		&$~~	1.00	$&$	-0.21	$&$	-0.19	$&$~~	0.59	$&$	-0.07	$&$~~	0.10	$	\\
$\alpha_{\CP}^{\rho_{\|}}$ 	&		&		&		&$~~	1.00	$&$~~	0.01	$&$	-0.04	$&$	-0.25	$&$	-0.09	$	\\
$\Delta2\beta^{\rm eff}_{\rm other}$	&		&		&		&		&$~~	1.00	$&$~~	0.00	$&$~~	0.26	$&$	-0.16	$	\\
$\Delta2\beta^{\rm eff}_{\rho_{\perp}}$	&		&		&		&		&		&$~~	1.00	$&$~~	0.39	$&$	-0.08	$	\\
$\Delta2\beta^{\rm eff}_{\rho_\|}$	&		&		&		&		&		&		&$~~	1.00	$&$	-0.10	$	\\
$2\beta^{\rm eff}_{\rho_0}$	&		&		&		&		&		&		&		&$~~	1.00	$	\\\hline
\end{tabular}
\def\arraystretch{1.0}
\end{center}
\end{table}

\begin{figure}[t]
\begin{center}
\includegraphics[width=0.65\textwidth]{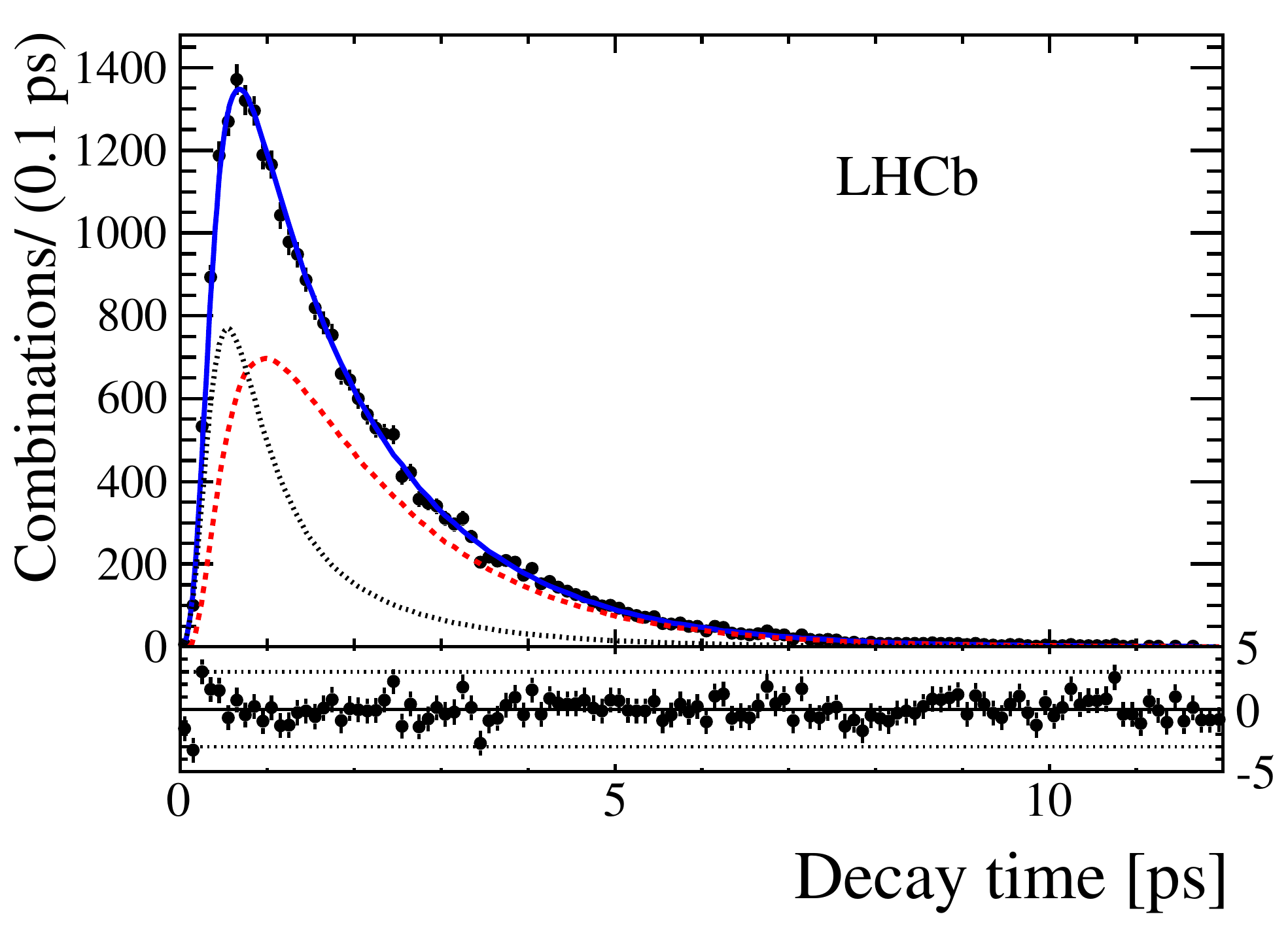}
\caption{Decay time distribution of $\BorBbar^0\to\jpsi \pi^+\pi^-$ candidates. The signal component is shown with a (red) dashed line, the background with a (black) dotted line, and the (blue) solid line represents the total. The lower plot shows the normalized residual distribution. }\label{finalfit}
\end{center}
\end{figure}

Table~\ref{tab:ff} lists the fit fractions and three transversity fractions of contributing resonances from  Fit 1, consistent with the results shown in the amplitude analysis~\cite{Aaij:2014siy}. For a $P$- or $D$-wave resonance, we report its total fit fraction by summing all three transversity components.
\begin{table}[b]
\centering
\caption{Fit and transversity fractions of contributing resonances from  Fit 1. Uncertainties are statistical only.  These results are presented only as a cross-check.}
\def\arraystretch{1.2}
\begin{tabular}{lcccc}\hline
 & & \multicolumn{3}{c}{Transversity fractions (\%)}\\
\raisebox{2.0ex}[0pt]{Component} & \raisebox{2.0ex}[0pt]{Fit fraction (\%)}& $0$ & $\|$ & $\perp$ \\\hline
$\rho(770)$ & $65.6\pm1.9$ & $56.7\pm1.8$& $23.5\pm1.5$& $19.8\pm1.7$\\
$f_0(500)$ & $20.1\pm0.7$ & 1 & 0 & 0\\
$f_2(1270)$ &$7.8\pm0.6$&$64\pm4$&$9\pm5$&$27\pm5$\\
$\omega(782)$&$0.64_{-0.13}^{+0.19}$& $44\pm14$&$53\pm14$&$3_{-3}^{+10}$\\
$\rho(1450)$&$9.0\pm1.8$& $47\pm11$&$39\pm12$&$14\pm8$\\
$\rho(1700)$&$3.1\pm0.7$&$29\pm12$&$42\pm15$&$29\pm15$\\\hline\end{tabular}\label{tab:ff}
\def\arraystretch{1.0}
\end{table}
This time-dependent analysis determines the phase difference between the \CP-odd component of $\rho(770)_\perp$ and the \CP-even component of $\rho(770)_0$ to be $(167\pm11)^\circ$ in Fit 1. This quantity is not accessible in the time-integrated amplitude analysis. Figure~\ref{finalfit} shows the decay time distribution superimposed with the fit projection.

The statistical significance of the \CP measurements are ascertained by fitting the data requiring that \CP-violating components are zero. We find that for the entire final state, this requirement changes  $-2$ times the logarithm of the likelihood  ($-2\ln {\cal{L}}$) by 28.6, corresponding to 4.4 standard deviations for four degrees of freedom (ndf), and for the $\rho(770)$ component only, the change is 24.0, corresponding to 4.5 standard deviations for two ndf. Here we only consider the statistical uncertainties.

\begin{table}[t]
\begin{center}
\caption{Results allowing for different \CP-violating effects for resonances other than the $\rho$ in an extension of Fit 1.}\label{dataextfit1}
\def\arraystretch{1.25}
\begin{tabular}{lc|lc}\hline
 \multicolumn{2}{c|}{$2\beta_i^{\rm eff}$ ($^\circ$)}& \multicolumn{2}{c}{$\alpha_{\CP}^i$ ($\times10^{-3}$)}\\\hline
$\rho$	&$	41.8	\pm	9.6$&	$\rho$	&$	~~~~2	\pm	39	$\\
	$f_0(500)-\rho$	&$	~~2.7	\pm	3.8$&	$f_0(500)$	&$	-58	\pm	46	$\\
$f_2(1270)-\rho$	&$	~~1.8	\pm	7.5$&	$f_2(1270)$	&$	~~~~\,9	\pm	63$\\
	other spin-1 $-\rho$	&$	~~~3.7\pm11.1$&	other spin-1	&$	~\,\,\,\,15	\pm	58 $\\\hline
\end{tabular}
\def\arraystretch{1.0}
\end{center}
\end{table}

We also perform a fit by extending Fit 1 to allow different \CP-violating effects in final states with either the $f_0(500)$, the $f_2(1270)$, or  spin-1 resonances. The results are shown in Table~\ref{dataextfit1}. We find that these fits all give consistent values of the \CP-violating parameters.

The systematic uncertainties evaluated for both fit configurations are summarized for the \CP-violating phases in Table~\ref{tab:sys1} and for the magnitudes of the  asymmetries in Table~\ref{tab:sys2}. They are small compared to the statistical ones.
The two largest contributions result from the resonance fit model and the resonance parameters.  Fit model uncertainties are determined by  adding an additional resonance to the default six-resonance model, either the $f_0(980)$, the $f_0(1500)$,  the $f_0(1700)$, or non-resonant $\pi^+\pi^-$,
replacing the $f_0(500)$ model by a Breit-Wigner function, and using the alternative Gounaris-Sakurai model shapes \cite{GS} for the various $\rho$ mesons. The largest variation among those changes is assigned as the systematic uncertainty for modelling. Including a non-resonant component gives the largest negative change on $2\beta^{\rm eff}$ for the $\rho$ and $\rho_0$ categories.

\begin{table}[b]
\begin{center}
\caption{Systematic uncertainties on \CP-violating phases $2\beta_{i}^{\rm eff}$ ($^\circ$). Statistical uncertainties are also shown.\label{tab:sys1}}
\begin{tabular}{lccc|ccc}\hline
Fit & \multicolumn{3}{c|}{Fit 1}&\multicolumn{3}{c}{Fit 2} \\
Sources	&	$\rho$ & other$-\rho$	& $\rho_0$ & $\rho_{\|}-\rho_0$ & $\rho_{\perp}-\rho_0$& other$-\rho_0$ \\\hline								
\rule{0pt}{2.5ex}Resonance model	&	$_{-	5.94	}^{+	1.85	}$&	$_{-	0.33	}^{+	0.51	}$&	$_{-	6.56	}^{+	1.99	}$&	$_{-	0.05	}^{+	1.35	}$&	$_{-	0.59	}^{+	1.50	}$&	$_{-	0.52	}^{+	0.68	}$	\\
Resonance parameters	&	$	\pm	1.21		$&	$	\pm	0.43		$&	$	\pm	1.35		$&	$	\pm	0.68		$&	$	\pm	0.57		$&	$	\pm	0.60		$	\\
Mass $\&$ angular acceptance	&	$	\pm	0.27		$&	$	\pm	0.05		$&	$	\pm	0.28		$&	$	\pm	0.21		$&	$	\pm	0.16		$&	$	\pm	0.05		$	\\
Angular acc. correlation	&	$	\pm	0.22		$&	$	\pm	0.03		$&	$	\pm	0.22		$&	$	\pm	0.21		$&	$	\pm	0.08		$&	$	\pm	0.03		$	\\
Decay time acceptance	&	$	\pm	0.05		$&	$	\pm	0.02		$&	$	\pm	0.06		$&	$	\pm	0.04		$&	$	\pm	0.04		$&	$	\pm	0.03		$	\\
Bkg.~mass $\&$ angular PDF	&	$	\pm	0.43		$&	$	\pm	0.09		$&	$	\pm	0.47		$&	$	\pm	0.22		$&	$	\pm	0.26		$&	$	\pm	0.11		$	\\
Bkg.~decay time PDF	&	$	\pm	0.14		$&	$	\pm	0.05		$&	$	\pm	0.12		$&	$	\pm	0.06		$&	$	\pm	0.08		$&	$	\pm	0.07		$	\\
Bkg.~model	&	$	\pm	0.49		$&	$	\pm	0.23		$&	$	\pm	0.15		$&	$	\pm	0.97		$&	$	\pm	0.38		$&	$	\pm	0.13		$	\\
Flavour Tagging	&	$	\pm	1.46		$&	$	\pm	0.03		$&	$	\pm	1.66		$&	$	\pm	0.44		$&	$	\pm	0.86		$&	$	\pm	0.01		$	\\
Production asymmetry	&	$	\pm	0.17		$&	$	\pm	0.50		$&	$	\pm	0.28		$&	$	\pm	0.09		$&	$	\pm	0.49		$&	$	\pm	0.42		$	\\\hline
\rule{0pt}{2.5ex}Total systematic uncertainty	&	$_{-	6.3	}^{+	2.8	}$&	$_{-	0.8	}^{+	0.9	}$&	$_{-	6.9	}^{+	3.0	}$&	$_{-	1.3	}^{+	1.9	}$&	$_{-	1.4	}^{+	2.0	}$&	$_{-	0.9	}^{+	1.0	}$	\\\hline
																																
Statistical uncertainty	&	$	\pm	9.6		$&	$	\pm	3.6		$&	$	\pm	10.2		$&	$	\pm	6.5		$&	$	\pm	7.2		$&	$	\pm	3.9		$	\\\hline
\end{tabular}
\end{center}
\end{table}

\begin{table}[bt]
\begin{center}
\caption{Systematic uncertainties for the magnitude of the asymmetries $\alpha_{\CP}^i$ ($\times10^{-3}$). Statistical uncertainties are also shown.\label{tab:sys2}}
\begin{tabular}{lccc|ccc}\hline
Fit & \multicolumn{3}{c|}{Fit 1}&\multicolumn{3}{c}{Fit 2}\\
Sources	&	$\rho$& other$-\rho$	& $\rho_0$ & $\rho_{\|}$ & $\rho_{\perp}$& other$-\rho_0$\\\hline																
\rule{0pt}{2.5ex}Resonance model	&	$_{-	0.0	}^{+	6.0	}$&	$_{-	11.4	}^{+	0.0	}$&	$_{-	0.0	}^{+	3.7	}$&	$_{-	2.7	}^{+	5.0	}$&	$_{-	0.0	}^{+	16.4	}$&	$_{-	11.0	}^{+	0.4	}$	\\
Resonance parameters	&	$	\pm	5.2		$&	$	\pm	6.1		$&	$	\pm	7.8		$&	$	\pm	3.1		$&	$	\pm	9.2		$&	$	\pm	7.3		$	\\
Mass $\&$ angular acceptance	&	$	\pm	0.6		$&	$	\pm	0.5		$&	$	\pm	0.8		$&	$	\pm	0.8		$&	$	\pm	1.6		$&	$	\pm	0.7		$	\\
Angular acc. correlation	&	$	\pm	0.2		$&	$	\pm	0.9		$&	$	\pm	0.2		$&	$	\pm	0.9		$&	$	\pm	0.6		$&	$	\pm	0.9		$	\\
Decay time acceptance	&	$	\pm	0.1		$&	$	\pm	0.1		$&	$	\pm	0.2		$&	$	\pm	0.3		$&	$	\pm	1.1		$&	$	\pm	0.1		$	\\
Bkg.~mass $\&$ angular PDF	&	$	\pm	0.9		$&	$	\pm	1.5		$&	$	\pm	0.8		$&	$	\pm	2.5		$&	$	\pm	4.6		$&	$	\pm	1.2		$	\\
Bkg.~decay time PDF	&	$	\pm	0.5		$&	$	\pm	0.4		$&	$	\pm	0.6		$&	$	\pm	0.5		$&	$	\pm	1.7		$&	$	\pm	0.4		$	\\
Bkg.~model	&	$	\pm	2.6		$&	$	\pm	2.9		$&	$	\pm	5.2		$&	$	\pm	3.5		$&	$	\pm	0.9		$&	$	\pm	4.6		$	\\
Flavour Tagging	&	$	\pm	2.8		$&	$	\pm	2.5		$&	$	\pm	0.5		$&	$	\pm	1.0		$&	$	\pm	10.7		$&	$	\pm	1.6		$	\\
Production asymmetry	&	$	\pm	3.0		$&	$	\pm	0.5		$&	$	\pm	2.5		$&	$	\pm	1.1		$&	$	\pm	0.4		$&	$	\pm	0.3		$	\\\hline
\rule{0pt}{2.5ex}Total systematic uncertainty	&	$_{-	7	}^{+	9	}$&	$_{-	14	}^{+~7	}$&	$_{-	10	}^{+	11	}$&	$_{-	6	}^{+	8	}$&	$_{-	15	}^{+	22	}$&	$_{-	14	}^{+~9	}$	\\\hline
Statistical uncertainty	&	$	\pm	28		$&	$	\pm	25		$&	$	\pm	34		$&	$	\pm	60		$&	$	\pm	109		$&	$	\pm	27		$	\\\hline

\end{tabular}
\end{center}
\end{table}

To evaluate the uncertainties due to the fixed parameters of resonances, we repeat the amplitude fit by varying the mass and width of all the resonances used in the six-resonance model within their errors one at a time, and add the changes in quadrature. To evaluate the systematic uncertainties due to the other fixed parameters including those in the decay time acceptance, the background decay time PDF, the $m(\pi^+\pi^-)$ distribution, the angular acceptance, and background mass PDF, the data fit is repeated by varying the fixed parameters from their nominal values according to the error matrix, one hundred times for each source. The matrix elements are determined using simulation, $\BorBbar^0\to\jpsi\KorKbar^{*0}$ data, and like-sign $\jpsi \pi^{\pm}\pi^{\pm}$ data. The r.m.s.\,of the fitted physics parameter of interest is taken as its uncertainty for each source.

The acceptance model for each of the three angles as a function of $\m$ is determined independently.
To evaluate the reliability of this method we parameterize the mass and angle efficiencies as a combination of Legendre polynomials and spherical harmonics that takes into account all correlations. The amplitude fit is repeated using the new acceptance parameterizations; changes are found to be small and taken as the systematic uncertainty.

In the nominal fit the background is divided into three sources: background to the $\rho^0$ component from $\BorBbar^0_{\!\!s}\to\jpsi\eta^\prime,\eta^\prime\to\rho^0\gamma$, reflection from $\BorBbar^0\to \jpsi \KorKbar^{*0}$ when the kaon is misidentified as a pion, and the remaining background. The latter includes the reflections from $\LorLbar^0_{\!\!b}\to\jpsi K^{\mp}\porpbar$ decays, where both the kaon and proton are misidentified, and combinatorial background.
 The dependence on $\m$ of the decay time distribution for this remaining background is modelled by using different decay time PDFs in different  $\m$ regions. We also change the background modelling by dividing the remaining background into separate combinatorial and $\Lb$ reflection components. The fit is repeated with the new background model,  and changes are taken as the systematic uncertainty.

The systematic uncertainty due to the tagging parameter calibration is given by the difference in quadrature
of the statistical uncertainties for each physics parameter between the nominal fit and an alternative fit where the tagging parameters are Gaussian constrained by their total uncertainties. The systematic uncertainty due to the asymmetry of   $\Bzb-\Bz$ meson production is estimated by varying the central value $A_{\rm P}=-0.0035\pm 0.0081$~\cite{Aaij:2014bba} by its uncertainty.

\section{Discussion of results and conclusions}

We compare the $\rho$-only Fit 1 result of $2\beta^{\jpsi\rho}=2\beta^{\rm eff}=(41.7\pm9.6_{-6.3}^{+2.8})^{\circ}$ with the Cabibbo-favoured $B$ to charmonium result, denoted $\jpsi\KS$. The measured difference is
\begin{equation}
\Delta 2\beta_f = 2\beta^{\jpsi\rho}-2\beta^{\jpsi\KS}=(-0.9\pm 9.7^{+2.8}_{-6.3})^{\circ}.
\end{equation}

\begin{figure}[b!]
  \begin{center}
  \vspace{-2mm}
     \includegraphics[width=0.72\textwidth]{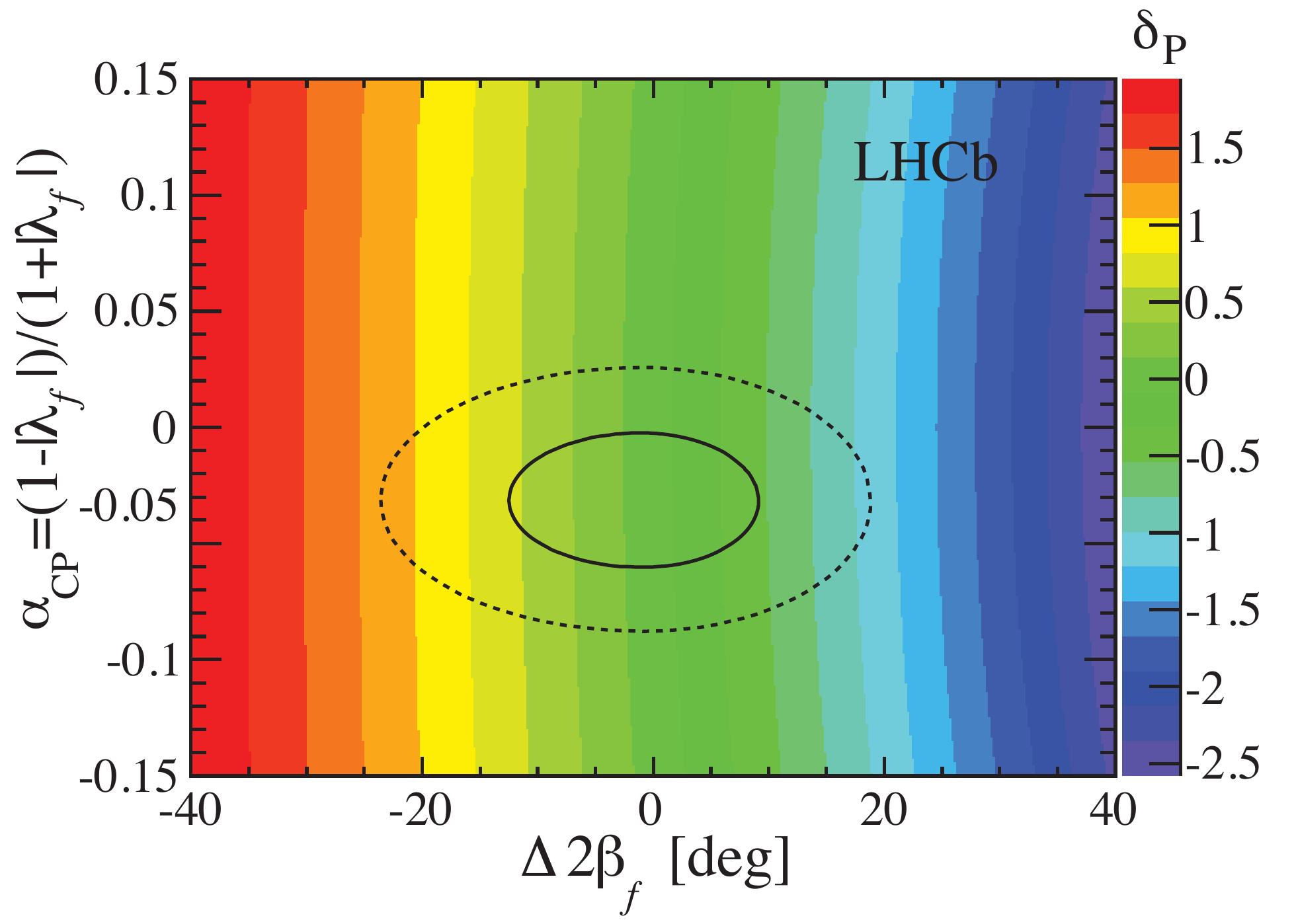}
     \vspace{-2mm}
    \caption{The magnitude of the penguin induced shift $\delta_P$ on the \CP-violating phase in favoured decays, assuming SU(3) flavour symmetry, shown in grey or colour scale in degrees, as a function of the measured difference $\Delta2\beta_f$ ($x$-axis) and $\alpha_{\CP}=\frac{1-|\lambda_f|}{1+|\lambda_f|}$ ($y$-axis).  Here we use fixed values for $\gamma=70^{\circ}$ and $\epsilon=0.0534$. The projected 68\% (solid) and 95\% (dashed) confidence levels on $\delta_P$ are shown by the egg-shaped contours.}  \label{Rel_sen}
  \end{center}
\end{figure}

Since the result is consistent with zero we determine limits on the magnitude of the \CP-violating phase shift due to a possible  penguin component in $b\to c\overline{c}s$ decays, $\delta_P$. The limit is evaluated using pseudo-experiments
by generating datasets with different values of $\alpha_{\CP}$, $2\beta^{\jpsi\rho}-2\beta^{\jpsi\KS}$, and $\gamma=(70.0_{-9.0}^{+7.7})^\circ$ \cite{HFAG} according to the measured uncertainties, including the correlation of $-0.01$ between $\alpha_{\CP}$ and $2\beta^{\jpsi\rho}$. Then $\delta_P$ for each dataset is calculated using Eq.~(\ref{EqRel}).
We find a Gaussian distribution with a 95\% confidence level (CL) interval of $[-1.05^\circ,1.18^\circ]$. This result is consistent with that obtained by projecting a contour of $\alpha_{\CP}$ and $2\beta^{\jpsi\rho}-2\beta^{\jpsi\KS}$ with regions proportional to the total uncertainties of the two physics variables as shown in Fig.~\ref{Rel_sen}.

\begin{figure}[b!]
  \begin{center}
  \vspace{-2mm}
     \includegraphics[width=0.64\textwidth]{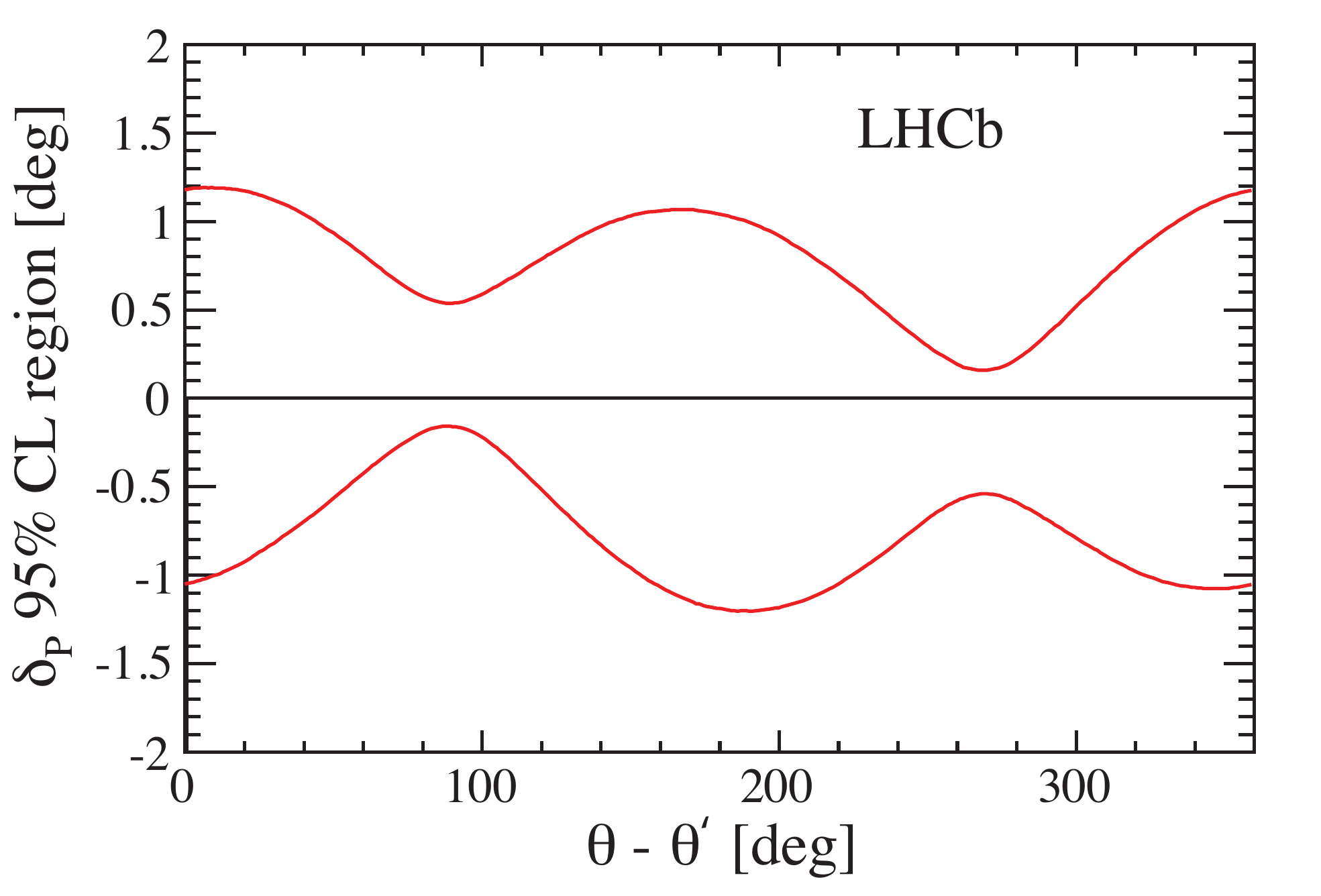}
    \caption{The limit on the penguin induced phase change $\delta_P$ as a function of the difference in the penguin amplitude strong phases in $b\to c\bar{c}s$ and $b\to c\bar{c}d$ transitions $\theta-\theta'$, for $a=a'$. }  \label{Ubreaking}
  \end{center}
\end{figure}

The two reactions $\BorBbar^0\to\jpsi\rho^0$ and $\BorBbar_s^0\to \jpsi\phi$ are related by SU(3) symmetry if we also assume that the difference between the $\phi$ being mostly a singlet state, and the $\rho^0$ an octet state causes negligible breaking.
Taking the magnitudes of the penguin amplitudes $a=a'$ and the strong phases $\theta=\theta'$ to be equal in $\BorBbar^0\to\jpsi\rho^0$ and $\BorBbar_{\!\!s}^0\to \jpsi\phi$ decays, and neglecting higher order diagrams \cite{Fleischer:1999sj}, we find $\delta_P=(0.05\pm 0.56)^{\circ}=0.9\pm 9.8$\, mrad. At  95\% CL, the penguin contribution in $\BorBbar_{\!\!s}^0\to \jpsi\phi$ decay is within the interval from $-1.05^\circ$ to +$1.18^\circ$. 
Relaxing these assumptions changes the limits on the possible penguin induced shift. Figure~\ref{Ubreaking} shows how $\delta_P$ varies as a function of $\theta-\theta'$, indicating that the 95\% CL limit on penguin pollution can increase to at most $\pm1.2^\circ$. The variation in $\delta_P$ is proportional to $a/a'$. Thus, when
changing $a/a'$ over the interval 0.5 to 1.5, the limit on the penguin shift at 95\% CL  varies between $\pm 0.9^{\circ}$ to $\pm 1.8^{\circ}$, even allowing for maximal breaking between $\theta'$ and $\theta$. 
It may be expected that the effect of penguin contributions in other decays, such as $\BorBbar^0\to\jpsi\KS$, should be limited to similar values, even if there is no strict flavour symmetry relating the mode to $\BorBbar^0\to\jpsi\rho^0$. Our limit
 is consistent with theoretical predictions \cite{Liu:2013nea,*Li:2006vq,*Jung:2012vd,*Ciuchini:2011kd,*Ciuchini:2005mg}.

\begin{figure}[t]
  \begin{center}
  \vspace{-12mm}
     \includegraphics[width=0.64\textwidth]{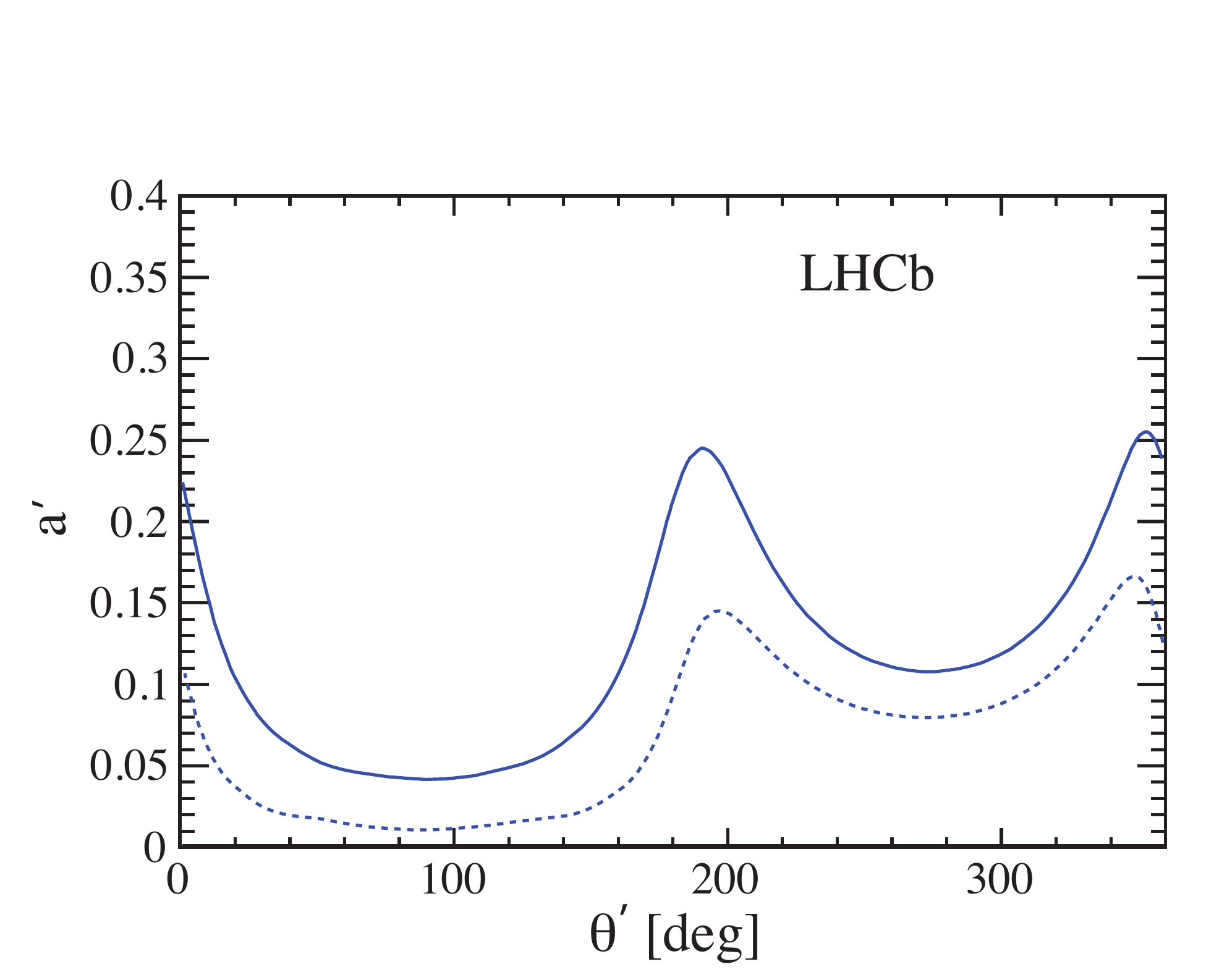}
     \vspace{-2mm}
    \caption{Contours corresponding to 68\% (dashed) and 95\% (solid) confidence levels for ndf of two, respectively, for the penguin amplitude parameters $a^\prime$ and $\theta^\prime$.}  \label{pen-con}
  \end{center}
\end{figure}

We also set limits on the strong decay amplitude. Figure~\ref{pen-con} shows the 68\% and 95\% confidence levels  contours 
for the penguin amplitude parameters of $a^\prime$ and $\theta^\prime$ with a $-2\ln {\cal{L}}$ change of 2.3 and 6 units,  for ndf equals two, including systematic uncertainties. They are obtained by converting the corresponding contours for $\alpha_{\CP}^{\jpsi\rho}$ and $\Delta2\beta_f$ using their relationship given in Eq.\,(\ref{pen1}). The uncertainty on the angle $\gamma=(70.0_{-9.0}^{+7.7})^\circ$ only introduces about a 0.2\% increase in the mean contour radius of $a^\prime$ versus $\theta^\prime$. The one-dimensional 68\% confidence level intervals are found by changing $-2\ln {\cal{L}}$ by one unit, giving  $a^\prime < 0.12$ and $\theta^\prime \in (190^\circ,355^\circ)$, or $a^\prime = 0.035_{-0.035}^{+0.082}$ and $\theta^\prime = (285_{-95}^{+69})^\circ$.

The decay $\BorBbar^0\to\jpsi\pi^0$ proceeds through a similar diagram to that shown in Fig.~\ref{feyn3}, and thus the \CP-violating parameters  $S$ and $C$ should be similar to those we find in $\BorBbar^0\to\jpsi\rho^0$.
These parameters are related to the parameter $\lambda_f$ via the relationships
\begin{equation}
S_f\equiv\frac{2{\cal I}m(\lambda_f)}{1+|\lambda_f|^2}=-2\eta_f\frac{|\lambda_f|\sin{2\beta^{\rm eff}_f}}{1+|\lambda_f|^2}, \quad {\rm and} \quad C_f\equiv \frac{1-|\lambda_f|^2}{1+|\lambda_f|^2},
\end{equation}
where we set the \CP eigenvalue $\eta_f=1$ to compare with the \CP-even mode $\BorBbar^0\to\jpsi\pi^0$.

Using $S_f$ and $C_f$ as fit parameters, we obtain from Fit 1 $S_{\jpsi\rho}=-0.66_{-0.12-0.03}^{+0.13+0.09}$ and $C_{\jpsi\rho}=-0.063\pm0.056^{+0.019}_{-0.014}$, with a correlation of $-0.01$. Table~\ref{cmp} shows the comparison of $S_f$ and $C_f$ from this measurement with that obtained from the Belle~\cite{Lee:2007wd} and BaBar~\cite{Aubert:2008bs} collaborations. Our measurements are in good agreement with the Belle results.
\begin{table}[htb]
\centering
\caption{Comparison of $S_f$ and $C_f$ between different measurements.}
\label{cmp}
\begin{tabular}{clccc}\hline
$f$& Experiment & $S_f$ & $C_f$ & Correlation \\\hline
\rule{0pt}{2.8ex}$\BorBbar^0\to\jpsi\rho^0$& LHCb & $-0.66_{-0.12-0.03}^{+0.13+0.09}$ & $-0.063\pm0.056^{+0.019}_{-0.014}$ & $-0.01$ (stat)\\
$\BorBbar^0\to\jpsi\pi^0$& Belle~\cite{Lee:2007wd} & $-0.65 \pm 0.21 \pm 0.05$  &	$-0.08 \pm 0.16 \pm 0.05$ &	$-0.10$ (stat)\\
$\BorBbar^0\to\jpsi\pi^0$& BaBar~\cite{Aubert:2008bs} & $-1.23 \pm 0.21 \pm 0.04$ 	& $-0.20 \pm 0.19 \pm 0.03$ & 	~~0.20 (stat)\\
\hline\end{tabular}
\end{table}

In conclusion, the measured value of the penguin contribution  is $\delta_P=(0.05\pm 0.56)^{\circ}=0.9\pm 9.8$\, mrad. Taking the maximum breaking in phase and a range of breaking $0.5<a/a'<1.5$ the uncertainty on $\delta_P$ becomes $\pm 18$\,mrad.  The measured value of $\phi_s$ currently has an uncertainty of about 35\,mrad, and the value of $2\beta$ of 1.5$^{\circ}$ or 26\,mrad \cite{HFAG}. Thus our limit is smaller than the current uncertainties, but will need to become more precise as the \CP-phase measurements improve.

\section*{Acknowledgements}
\noindent We express our gratitude to our colleagues in the CERN
accelerator departments for the excellent performance of the LHC. We
thank the technical and administrative staff at the LHCb
institutes. We acknowledge support from CERN and from the national
agencies: CAPES, CNPq, FAPERJ and FINEP (Brazil); NSFC (China);
CNRS/IN2P3 (France); BMBF, DFG, HGF and MPG (Germany); SFI (Ireland); INFN (Italy); 
FOM and NWO (The Netherlands); MNiSW and NCN (Poland); MEN/IFA (Romania); 
MinES and FANO (Russia); MinECo (Spain); SNSF and SER (Switzerland); 
NASU (Ukraine); STFC (United Kingdom); NSF (USA).
The Tier1 computing centres are supported by IN2P3 (France), KIT and BMBF 
(Germany), INFN (Italy), NWO and SURF (The Netherlands), PIC (Spain), GridPP 
(United Kingdom).
We are indebted to the communities behind the multiple open 
source software packages on which we depend. We are also thankful for the 
computing resources and the access to software R\&D tools provided by Yandex LLC (Russia).
Individual groups or members have received support from 
EPLANET, Marie Sk\l{}odowska-Curie Actions and ERC (European Union), 
Conseil g\'{e}n\'{e}ral de Haute-Savoie, Labex ENIGMASS and OCEVU, 
R\'{e}gion Auvergne (France), RFBR (Russia), XuntaGal and GENCAT (Spain), Royal Society and Royal
Commission for the Exhibition of 1851 (United Kingdom).

\clearpage
\newpage
\newpage
\addcontentsline{toc}{section}{References}
\setboolean{inbibliography}{true}
\ifx\mcitethebibliography\mciteundefinedmacro
\PackageError{LHCb.bst}{mciteplus.sty has not been loaded}
{This bibstyle requires the use of the mciteplus package.}\fi
\providecommand{\href}[2]{#2}

\newpage


\centerline{\large\bf LHCb collaboration}
\begin{flushleft}
\small
R.~Aaij$^{41}$, 
B.~Adeva$^{37}$, 
M.~Adinolfi$^{46}$, 
A.~Affolder$^{52}$, 
Z.~Ajaltouni$^{5}$, 
S.~Akar$^{6}$, 
J.~Albrecht$^{9}$, 
F.~Alessio$^{38}$, 
M.~Alexander$^{51}$, 
S.~Ali$^{41}$
% ...

{\footnotesize \it
% ...
$ ^{5}$Clermont Universit\'{e}, Universit\'{e} Blaise Pascal, CNRS/IN2P3, LPC, Clermont-Ferrand, France\\
$ ^{6}$CPPM, Aix-Marseille Universit\'{e}, CNRS/IN2P3, Marseille, France\\
% ...
$ ^{9}$Fakult\"{a}t Physik, Technische Universit\"{a}t Dortmund, Dortmund, Germany\\
% ...
$ ^{37}$Universidad de Santiago de Compostela, Santiago de Compostela, Spain\\
$ ^{38}$European Organization for Nuclear Research (CERN), Geneva, Switzerland\\
% ...
$ ^{41}$Nikhef National Institute for Subatomic Physics, Amsterdam, The Netherlands\\
% ...
$ ^{46}$H.H. Wills Physics Laboratory, University of Bristol, Bristol, United Kingdom\\
% ...
$ ^{51}$School of Physics and Astronomy, University of Glasgow, Glasgow, United Kingdom\\
$ ^{52}$Oliver Lodge Laboratory, University of Liverpool, Liverpool, United Kingdom\\
% ...
}
\end{flushleft}

\end{document}